\documentclass[letterpaper, 10 pt, conference]{ieeeconf}  

\IEEEoverridecommandlockouts                              
\usepackage{mathrsfs}
\usepackage{subfigure}
\usepackage{lipsum}
\usepackage{multicol}
\usepackage{graphicx}
\usepackage{multirow}
\usepackage{threeparttable}
\usepackage{amsmath}
\usepackage{caption}

\title{\LARGE \bf
{Deep Reinforcement Learning of Cell Movement in the Early Stage of {\em C. elegans} Embryogenesis}
}

\author{Zi Wang$^{1}$, Dali Wang$^{1,3,*}$, Chengcheng Li$^{1}$, Yichi Xu$^{2}$, Husheng Li$^{1}$ and Zhirong Bao$^{2,*}$
\thanks{$^{1}$Zi Wang, Dali Wang, Chengcheng Li and Husheng Li are with the Department of Electrical Engineering and Computer Science, University of Tennessee, Knoxville, TN 37934, USA. \{zwang84,dwang7,cli42,hli31\}@utk.edu.}%
\thanks{$^{2}$Yichi Xu and Zhirong Bao are with the Developmental Biology Program, Sloan-Kettering Institute, New York, NY 10065, USA. \{xuy2,baoz\}@mskcc.org.}%
\thanks{$^{3}$Dali Wang is also with the Environmental Science Division, Oak Ridge National Laboratory, Oak Ridge, TN 37831, USA.}
\thanks{$^{*}$Dali Wang and Zhirong Bao are the corresponding authors.}
}

\begin{document}

\maketitle
\thispagestyle{empty}
\pagestyle{empty}

\begin{abstract}

Cell movement in the early phase of {\em C. elegans} development is regulated by a highly complex process in which a set of rules and connections are formulated at distinct scales. Previous efforts have demonstrated that agent-based, multi-scale modeling systems can integrate physical and biological rules and provide new avenues to study developmental systems. However, the application of these systems to model cell movement is still challenging and requires a comprehensive understanding of regulation networks at the right scales.  Recent developments in deep learning and reinforcement learning provide an unprecedented opportunity to explore cell movement using 3D time-lapse microscopy images.  \\
We present a deep reinforcement learning approach within an agent-based modeling system to characterize cell movement in the embryonic development of {\em C. elegans}. Our modeling system captures the complexity of cell movement patterns in the embryo and overcomes the local optimization problem encountered by traditional rule-based, agent-based modeling that uses greedy algorithms. We tested our model with two real developmental processes: {\em the anterior movement of the Cpaaa cell via intercalation} and {\em the rearrangement of the superficial left-right asymmetry}. In the first case, the model results suggested that Cpaaa's intercalation is an active directional cell movement caused by the continuous effects from a longer distance, as opposed to a passive movement caused by neighbor cell movements. This is because the learning-based simulation found that a passive movement model could not lead Cpaaa to the predefined destination. In the second case, a leader-follower mechanism well explained the collective cell movement pattern in the asymmetry rearrangement. These results showed that our approach to introduce deep reinforcement learning into agent-based modeling can test regulatory mechanisms by exploring cell migration paths in a reverse engineering perspective. This model opens new doors to explore the large datasets generated by live imaging.
\end{abstract}

\section{Introduction}

Recent developments in cutting-edge live microscopy and image analysis provide an unprecedented opportunity to systematically investigate individual cells' dynamics and quantify cellular behaviors over extended period of time.  Systematic single-cell analysis of \emph{C. elegans} has led to the highly desired quantitative measurement of cellular behaviors \cite{murray2012multidimensional,du2014novo,du2015regulatory}. Based on 3D time-lapse imaging, the entire cell lineage can be automatically traced, and quantitative measurements can be made on every cell to characterize its developmental behaviors \cite{schnabel1997assessing,hench2009spatio,giurumescu2012quantitative,kyoda2012wddd}. These massive recordings, which contain hundreds to thousands of cells over hours to days of development, provide a unique opportunity for cellular-level systems behavior recognition as well as simulation-based hypothesis testing.

Agent-based modeling (ABM) is a powerful approach to analyze complex tissues and developmental processes \cite{setty2012multi,olivares2016virtual,hoehme2010cell}.  In our previous effort, an observation-driven, agent-based modeling and analysis framework was developed to incorporate large amounts of observational/phenomenological data to model the individual cell behaviors with straightforward interpolations from 3D time-lapse images \cite{wang2016observation,wang2017visualization}. With the ultimate goal being to model individual cell behaviors with regulatory mechanisms, tremendous challenges still remain to deal with the scenarios where regulatory mechanisms lag data collection and potential mechanistic insights need to be examined against complex phenomena.  

Directional cell movement is critical in many physiological processes during {\em C. elegans} development, including morphogenesis, structure restoration, and nervous system formation. It is known that, in these processes, cell movements can be guided by gradients of various chemical signals, physical interactions at the cell-substrate interface and other mechanisms \cite{lee2003mechanisms,shook2003mechanisms,lo2000cell}. It remains an open and interesting challenge as to what and how one could learn about the rules and mechanisms of cell movement from the movement tracks recorded in live imaging. 

This paper presents a new approach to study cell movement by adopting deep reinforcement learning approaches within an agent-based modeling framework. Deep reinforcement learning is good at dealing with high-dimensional inputs and can optimize complex policies over primitive actions \cite{mnih2013playing}, which naturally aligns with the complex cell movement patterns occurred during {\em C. elegans} embryogenesis. Even in some biological scenarios where regulatory mechanisms are not completely studied, deep neural networks can be adopted to characterize the cell movement within an embryonic system. The neural network takes information from 3D time-lapse images as direct inputs, and the output is the cell's movement action optimized under a collection of regulatory rules. Since deep reinforcement learning can optimize the cell migration path over considerable temporal and spatial spans in a global perspective, it overcomes the local optimization problem encountered by traditional rule-based, agent-based modeling that uses greedy algorithms.

We tested our model through two representative scenarios during \emph{C. elegans} embryogenesis: {\em the anterior movement of Cpaaa via intercalation} and {\em the rearrangement of the superficial left-right asymmetry}. In the first case, we proposed two hypotheses for the intercalation of Cpaaa, and simulation results indicated that Cpaaa experienced an active directional movement towards the anterior, which is caused by the continuous effects from a longer distance, rather than a passive process in which it is squeezed to the target location by its neighbors' movements. In the second case, the frequently occurring "leader-follower" mechanism was also supported by the simulation results of the asymmetry rearrangement. In summary, this framework presents a reverse engineering perspective to investigate regulatory mechanisms behind a certain developmental process: By formulating the reward functions as the representation of regulatory mechanisms, different hypotheses can be tested via reinforcement learning procedures. By comparing the extent of similarities between the simulation cell migration paths and the observation data, such hypotheses can either be supported or rejected, which can facilitate new explanations of certain cell movement behaviors. The model can also be used to study cell migration paths in {\em C. elegans} mutants or other metazoan embryo/tissue systems when related data are given.

\section{Modeling Approach}

In our modeling framework, an individual cell is modeled as an agent that contains a variety of information on its fate, size, division time, and group information. For a wild-type {\em C. elegans} simulation, the cell fate and division information can be directly derived from predefined observation datasets. For more complicated cases that involve gene mutation and manipulation, the developmental landscape can be incorporated for the purpose of modeling \cite{du2015regulatory}. More detailed design information on the agent-based model can be found in \cite{wang2016observation}. In this study, the cellular movements are treated as results of inherited and genetically controlled behaviors regulated by inter- or intracellular signals, and these cell movements are also constricted by the neighbor cells and the eggshell. 

We further assume that the migration path of an individual cell is the optimal path that a cell can use to migrate under a collection of regulation networks and/or constraints within a physical environment. Then we can transform the cell movement problem into a neural network construction and learning problem using observational and/or predefined rules. Therefore, neural networks can be constructed inside each cell to represent its behaviors, and the reinforcement learning method can be used to train the neural networks from 3D time-lapse imaging (with information on locations of cells, their neighbor lists, and other cell interactions after automated cell lineage tracing \cite{bao2006automated}). After training, the neural networks can determine a feasible and optimal cell migration path in a dynamic embryonic system, but the migration path is still controlled and constrained by the underlying regulation networks and the physical environment. 

While the regulation networks can be defined at cellular, group, tissue, or even embryonic levels, only the individual cell movement and group movement are examined and modeled in this study. 

\subsection{Individual Cell Movements}
\label{sec:individual}
Two basic kinds of individual cell movements are investigated. The first movement pattern is directional movement, in which the regulation network presents strong signals (such as morphogen gradient or planar cell polarity \cite{heisenberg2013forces}) and results in directional individual cell movements. The second type of cell movement, defined as passive cell movement, represents the scenarios in which no explicit movement patterns are observed when the signals from regulation networks are weak or canceled out.

\subsubsection{Directional cell movement}
At this stage, with strong regulation signals from regulation networks, cell movement is mainly controlled by the potential destination and physical pressures from neighbor cells or the eggshell. The destination of cell movement can be defined as a spatial location or region within the embryonic system when regulatory mechanisms are not well studied, or it can be defined as a location next to a specific cell. 

\subsubsection{Passive cell movement}
At this stage, without strong overall regulation mechanisms, cell movement is mainly controlled by the physical pressures between neighbor cells or the eggshell. Therefore, it is defined as passive cell movement with a high level of randomness. 

\subsection{Collective Cell Migration}

In a \emph{C. elegans} embryonic system, individual cells can also be a part of functional group with group-specific communication and regulation mechanisms. In collective cell migration, all the cell movements are directional. However, depending on the role of cell movement, the cells in collective migration can be further categorized as leading cells and following cells. 

\section{Methods}
\subsection{ABM Framework}

\begin{figure}[tpb]
\centering
\includegraphics[width=0.45\textwidth]{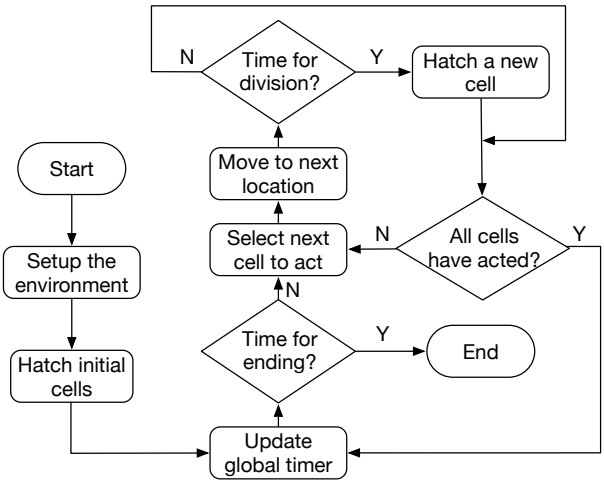}
\caption{The ABM framework. Cells move at each time step based on the output of the neural network (intelligent cell) or reading the observed locations (dumb cells). After a cell's movement, if it is at the right time for division, a new cell is hatched. Such a process repeats until the end of the simulation.}
\label{fig:abm}
\end{figure}

An ABM platform was adopted to present fundamental cell behaviors, including cell fate, division, and migration for a wild-type {\em C. elegans} in which all cell fates are predefined. The framework, which retains two fundamental characteristics (cell movement and division) for \emph{C. elegans} early embryogenesis is illustrated in Fig. \ref{fig:abm}. We use the terminologies ``intelligent cell'' and ``dumb cell'' to represent the cell that learns its migration path, and those move based on the observation dataset, respectively. At each time step, each cell first moves to its next location determined by either the output action from the neural network (if the cell is an ``intelligent cell'') or the observation data (if the cell is a ``dumb cell''). After that, if it is at the right time for division, a new cell is hatched. A global timer is updated when all the cells have acted at a single time step, and such a loop repeats until the end of the process.

\subsection{Cell Movement via Deep Q-network}
\begin{figure}[tpb]
\centering
\includegraphics[width=0.45\textwidth]{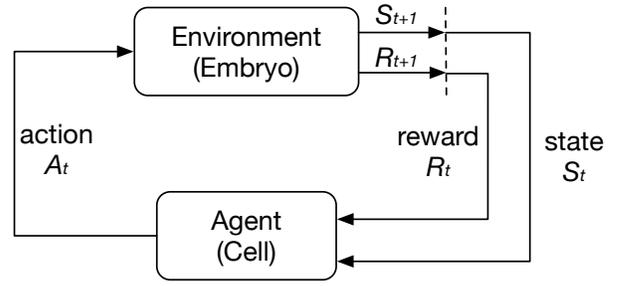}
\caption{The reinforcement learning framework. A cell interacts with the embryo. At each time step, the cell receives a state $S_t$, selects an action $A_t$, gets a reward $R_t$ and enters the next state $S_{t+1}$. The cell's objective is to maximize the total rewards received.}
\label{fig:rl}
\end{figure}

As mentioned in the Modeling Approach section, cell movement has been modeled as a reinforcement learning process \cite{sutton1998reinforcement} in which an agent (cell) interacts with the environment (other cells in the embryo and the eggshell) to achieve predefined goals. In an individual cell movement case, an intelligent cell always tends to seek the optimal migration path towards its destination based on the regulatory rules. At each discrete time step $t$, the cell senses its environmental state $S_t \in \mathcal{S}$ from the embryo and selects an action $A_t \in \mathcal{A}$, where the set of $\mathcal{A}$ includes the candidate actions at that state. The embryo returns a numerical reward $R_{t} \in \mathcal{R}$ to the cell as an evaluation of that action based on the state. Finally, the cell enters the next state $S_{t+1}$ and repeats the process until a terminal condition is triggered. The intelligent cell's objective is to maximize the overall rewards collected during the process. The whole process is demonstrated in Fig. \ref{fig:rl}.

Traditionally, tabular-based Q-learning approaches were largely used for reinforcement learning tasks with modest amounts of input states. However, a dynamic agent-based embryogenesis model usually contains hundreds of cells that act at high temporal and spatial resolutions. Millions of different states are generated during a single embryogenesis process, which cannot be handled by traditional tabular-based Q-learning algorithms. Furthermore, a traditional Q-learning algorithm requires large computational resources and can not be tightly integrated within an agent-based modeling framework for large-scale simulations with high-dimensional inputs. Recent breakthroughs in reinforcement learning that incorporate deep neural networks as mapping functions allow us to feed in high-dimension states and obtain the corresponding Q-values that indicate a cell's next movement \cite{mnih2013playing,mnih2015human}. Such a deep Q-network (DQN) outperforms most of the previous reinforcement learning algorithms.

\subsubsection{Framework}
\begin{figure}[tpb]
\centering
\includegraphics[width=0.45\textwidth]{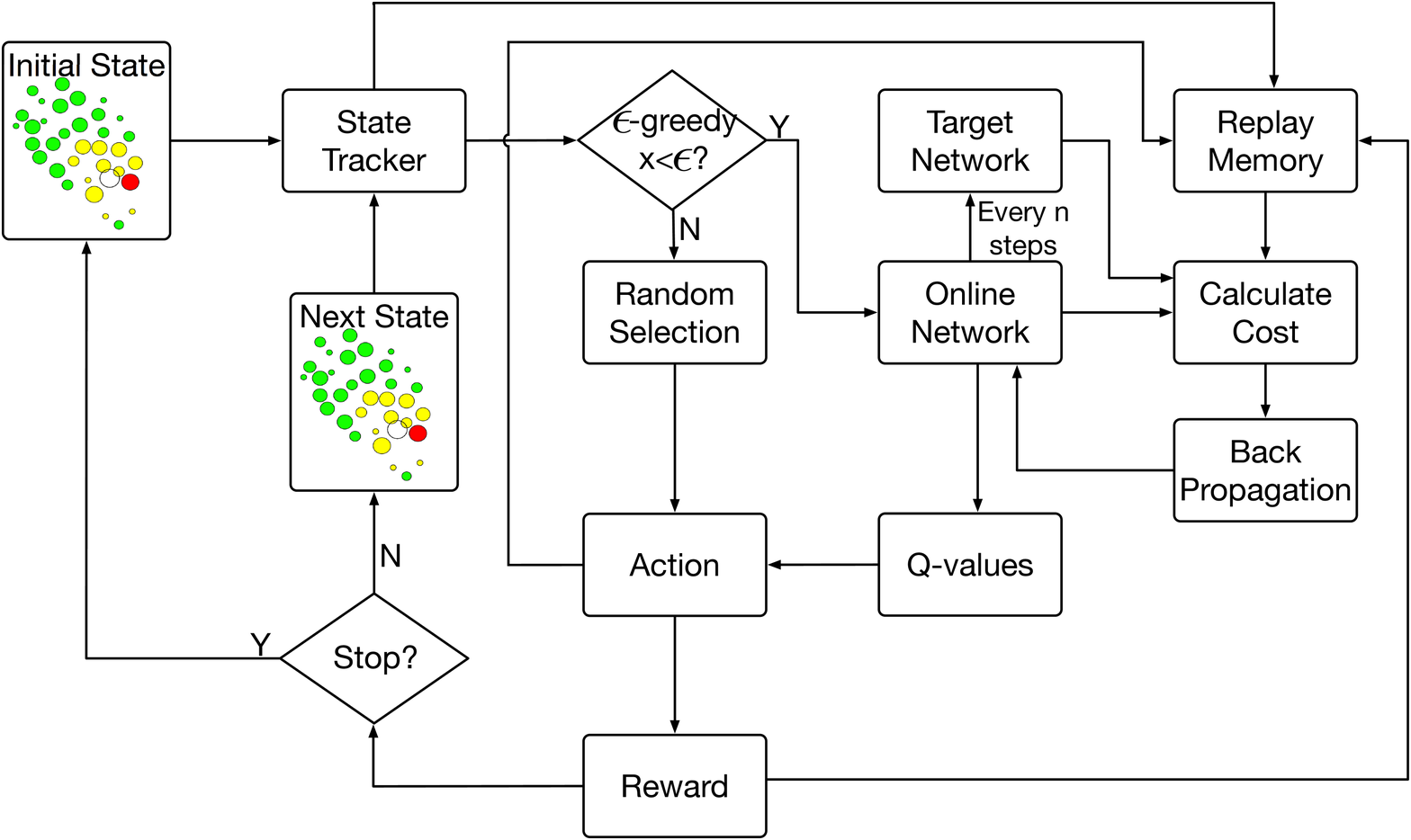}
\caption{The deep Q-network framework for cell movement, which cotnains a cell migration loop and a network learning loop. The intelligent cell's movement is selected via the $\epsilon$-greedy mechanism, from either a random sampling of all the possible actions or the output of the neural network. Then it gets a reward, moves to the next location, and repeats this process. The samples generated from the cell migration loop are used to update the parameters of the neural network via backpropagation. Experience replay and target network are implemented to improve the performance.}
\label{fig:dqn}
\end{figure}

We implemented a DQN customized for cell movement modeling. It contains two main loops: a cell migration loop and a network training loop (Fig. \ref{fig:dqn}). At each time step in the cell migration loop, a state tracker is used for collecting the input state as a representation of the environmental conditions (details in Section \ref{sec:input_states}). An $\epsilon$-greedy strategy is implemented to balance the exploration and exploitation. Specifically, $\epsilon$ is a hyperparameter in $[0,1)$. A random number $x$ is sampled from a uniform distribution $U(0,1)$ each time before the selection of an action. If $x \in [\epsilon,1)$, the intelligent selects a random action, obtains a reward and moves to the next location. Otherwise, the movement action is calculated by feeding the input state to the neural network. Such a process repeats until a terminal condition is triggered. For the training loop, the DQN is established based on traditional Q-learning algorithms. Rather than searching a Q-table to find the maximal value of $Q(S_{t},A_t)$, Q-values are obtained through a neural network parameterized by a set of weights $\theta$. The training samples are the tuples $(S_t,A_t,R_t,S_{t+1})$ gathered from the migration loop. The update process (Eq. (\ref{eq:dqn})) can be achieved by minimizing the loss function $\mathcal{L}$ (Eq. (\ref{eq:dqnloss})) and backpropagating the loss through the whole neural network to update $\theta$ by $\theta_{t+1}=\theta_t-\alpha\nabla_\theta\mathcal{L}(\theta_t)$ \cite{egorov2016multi}. Therefore, the intelligent cell will gradually select better actions as the training process proceeds.

\begin{equation}
\begin{aligned}
Q(S_t,&A_t|\theta_t) \leftarrow Q(S_t,A_t|\theta_t)\\
&+\alpha\left[R_t + \gamma\max_aQ(S_{t+1},A_{t+1}|\theta_t) - Q(S_t,A_t|\theta_t)\right],
\end{aligned}
\label{eq:dqn}
\end{equation}

\begin{equation}
\mathcal{L}(S_t,A_t|\theta_t) = \left[R_t + \gamma\max_aQ(S_{t+1},A_{t+1}|\theta_t) - Q(S_t,A_t|\theta_t)\right]^2,
\label{eq:dqnloss}
\end{equation}
where $\alpha$ is the learning rate and $\gamma \in (0,1)$ is the discount factor, which determines the present value of future rewards \cite{sutton1998reinforcement}.

In order to improve the system's performance, we utilized two mechanisms, i.e., experience replay \cite{mnih2013playing} and target network \cite{mnih2015human}, in the framework. Experience replay cuts off the correlation (which is one of the sources of instabilities) between samples by storing the movement tuples $(S_t, A_t, R_t, S_{t+1})$ in a replay memory and sampling them randomly during the training process. This is because the capacity of the replay buffer is much larger than the number of samples generated in a single process (from the beginning to a terminal state), and the randomly selected samples for training at each time will come from various processes, which are much less related with each other than those consecutive samples from a single process. In a DQN with a single neural network, the target for gradient descent is always shifting as $\theta$ is updated at each time step. Therefore, rather than calculating the future maximal expected reward $\max_aQ(S_{t+1},A_{t+1}|\theta_t)$ and updating the weights in a single neural network, a target network, which has the same architecture as the original network (called the online network in the new scenario) but parameterized with $\theta_t^-$, was implemented for the calculation of $\max_aQ(S_{t+1},A_{t+1}|\theta_t^-)$. The weights $\theta_t^-$ remains unchanged for all $n$ iterations until they are updated with $\theta_t$ from the online network. This mechanism reduces the oscillations and improve the stabilities of the framework. The improved process is represented in Eq. (\ref{eq:dqn_target}).

\begin{equation}
\begin{aligned}
Q(S_t,&A_t|\theta_t) \leftarrow Q(S_t,A_t|\theta_t)\\
&+\alpha\left[R_t + \gamma\max_aQ(S_{t+1},A_{t+1}|\theta_t^-) - Q(S_t,A_t|\theta_t)\right]
\end{aligned}
\label{eq:dqn_target}
\end{equation}

The neural network, which is fed with the environmental state and outputs a Q-value for each action, contains three hidden layers, with 512, 1024, and 1024 nodes, respectively. The Rectified Linear Units (ReLU) was implemented as the activation function after all the hidden layers except for the output layer. The details of the hyperparameter selection can be found in the Supplementary Material S1.1.

\subsubsection{Regulatory mechanisms and reward settings}
In the reinforcement learning scenario, the regulatory mechanisms that guide cell movements can be transformed to reward functions as an evaluation of how well a cell moves during a certain period of time based on those mechanisms. For the physical constraints of the cell movement, we defined the following two rules:

\begin{itemize}
\item \emph{Collision}: Cells cannot squeeze too much with each other. The closer two cells are, the larger penalty (negative reward) they receive.

\item \emph{Boundary}: Cells cannot break through the eggshell. The closer the cell is to the eggshell, the larger penalty (negative reward) it receives.
\end{itemize}

For both of the above rules, as a threshold of distance is reached, a terminal condition is triggered and the process ends and restarts. For the directional cell movement, an explicit destination is given as a simplified third rule when other regulatory mechanisms are missing:

\begin{itemize}
\item \emph{Destination}: A cell always seeks the optimal path towards its target location.
\end{itemize}

This rule can be replaced as more specific regulatory mechanisms are discovered (e.g., following a leading cell or becoming the neighbor of a certain cell), or new hypotheses are formulated. Details of the reward settings are illustrated in Section 4 and Supplementary Material S1.2.

\subsection{Behaviors of the Dumb Cells}
The automated cell lineage tracing technology was utilized to obtain the information of cells' identities and locations from 3D time-lapse microscopy images. These information were used to model the non-intelligent cells' (dumb cells') movement. Because the temporal resolution of our observation data is one minute, and an ABM simulation often requires a much smaller tick interval, a linear interpolation was implemented between two consecutive samples to calculate the next locations of these cells. Additionally, we added a random noise for each movement by sampling it from a normal distribution whose mean value and standard deviation were averaged from the locations of the cells of 50 wild-type \emph{C. elegans} embryos \cite{moore2013systematic}.

\subsection{Behaviors of the Intelligent Cell}
For the intelligent cell, an $\epsilon$-greedy strategy was implemented, which makes it not only act based on past experiences to maximize the accumulated rewards most of the time but also gives it a small chance to randomly explore unknown states. Usually, the value of $\epsilon$ is set to increase (the probability of random exploration decreases) as the training process proceeds. This is because the demands of exploration narrows down as the intelligent cell moves towards the destination. The selection of $\epsilon$ varies from case to case and the details are demonstrated in the Supplementary Material S1.1. In the following sub-sections, we give a description of the settings of the intelligent cell's input states and output actions.

\subsubsection{Input states}
\label{sec:input_states}
Representing the input state accurately and efficiently is a key issue for the deep reinforcement learning framework of cell movement. Besides the location of the intelligent cell, which is indispensable, an intuitive assumption is that its neighbors, which represent the environment, should be incorporated to form the input state. We implemented a neighbor determination model (which takes a set of features of two cells, such as the distance between them, their radii, etc., and determines whether they are neighbors with each other with machine learning algorithms) \cite{wang2017cell} in a conservative manner for this purpose. Specifically, we extracted a number of candidate cells that might influence the intelligent cell with a relatively loose condition, so that more cells would be selected to guarantee that the input state is sufficiently represented. This was done by running the agent-based model in a non-reinforcement learning mode (all cells move based on the observation data) and recording the neighbors of the intelligent cell at each time step. Finally, we combined the locations of all these cells (selected accumulatively in the whole process) in a fixed order as the input for the neural network.

\subsubsection{Output actions}
It is intuitive to give the intelligent cell as many candidates of actions as possible (or a continuous action space) so that it can make the most eligible choice during the simulation. The diversity of the action includes different speeds and directions. However, the number of output nodes grows exponentially as we take looser strategies to select the action. Based on our extensive experiments, we discovered that an enumeration of eight directions of action, with $45^{\circ}$ between each of them, is good enough for this scenario. Moreover, we fixed the speed based on an estimation of the average movement speed during the embryogenesis, which was measured from the observation data.

\begin{figure}[!tpb]
\begin{minipage}{1\columnwidth}
\centering
\includegraphics[width=0.95\textwidth]{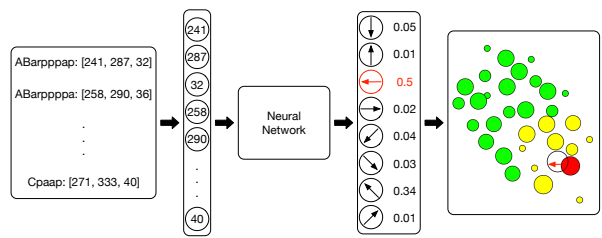}
\end{minipage}
\caption{An example of a specific evaluation step for a single action. A list of cells are pre-selected as the state cells via the cell neighbor determination model. Their locations are concatenated and sent to the neural network, and the output action with the maximal probability is selected as the intelligent cell's next movement.}
\label{fig:example}
\end{figure}

Finally, we give an example of a specific evaluation step for a single action selection process (Fig. \ref{fig:example}). We collect all the locations of the selected cells by the neighbor determination model, concatenate them to form a vector in a fixed order, and feed it into the neural network. The output of the neural network are the Q-values (i.e., a probability for selecting each action). The action that corresponds to the maximal probability (or a random action as the $\epsilon$-greedy suggested) is selected as the intelligent cell's next movement.

\subsection{Computational Environment and Platform}
The agent-based model was implemented with Mesa, which is an ABM framework in Python 3+. We used Python's GUI package Tkinter for the purpose of visualization. The cell movement behavior model was built with 3D coordinates, and certain slice of the whole embryo was visualized in a 2D manner to illustrate where emergent behaviors specifically happen. We used Pytorch to achieve reinforcement learning algorithms with the advantage of GPU acceleration during the training process. The reinforcement learning architecture was integrated as part of the agent-based model. All the computations were executed in a {DELL\textregistered} Precision workstation, configured with a 3.6 GHz 4-core {Intel\textregistered} {Xeon\textregistered} CPU, 64 GB main memory, and a 16-GB {NVIDIA\textregistered} {Quadro\textregistered} P5000 GPU.

\subsection{Model Setup}
Live 3D time-lapse images of \emph{C. elegans} embryogenesis data were used to study cell movement. Cell lineage \cite{sulston1983embryonic} was traced by Starrynite II \cite{santella2014semi} and manually corrected in Acetree \cite{boyle2006acetree}. Acetree was also used to visualize the observation data. Detailed information on live imaging can be found in the Supplementary Material S2. 

Two special {\em C. elegans} biological phenomena, {\em the intercalation of Cpaaa} and {\em left-right asymmetry rearrangement}, were investigated.  The first case is a remarkable process during {\em C. elegans} early morphogenesis of dorsal hypodermis. Cpaaa is born at the dorsal posterior. About 10 minutes later after its birth, Cpaaa moves towards the anterior and intercalates into two branches of ABarp cells, which will give rise to left and right seam cells, respectively. The intercalation of Cpaaa is consistent among wild-type embryos. It leads to the bifurcation of ABarp cells and the correct positioning of seam cells. The second case is {\em left-right asymmetry rearrangement}. It is a significant development scenario:  At the 4-cell stage, the left-right symmetry is broken after the skew of ABa/ABp spindle. The right cell  ABpr is positioned more posterior than the left cell ABpl. At the AB64 (64 AB cells, 88 total cells) stage, the movement of ABpl and ABpr cells start to restore the spatial symmetry, i.e., ABpl cells move towards the posterior and ABpr cells move towards the anterior. By 350-cell stage, ABpl and ABpr cells are again in symmetry on the AP axis. This asymmetry rearrangement achieves a superficially symmetric body plan \cite{pohl2010chiral}.

The embryo is considered to be an ellipsoid for the volume estimation. The mounting technique aligns the DV axis in the embryo with the z-axis of the data \cite{bao2006automated,bao2011mounting}, and the lengths of the other two axes (AP and LR) are obtained by finding the minimum and maximum cell positions along them \cite{moore2013systematic}. For the estimation of the cell radius, the ratio of the cell volume to the entire embryo is determined based on its identity. Then, the radius is estimated by considering a cell as a sphere \cite{wang2017cell}.

We utilized linear functions to define the rewards in our simulations. Specifically, for the \emph{Collision} rule, a penalty  (negative reward) is exerted as the distance between two cells reached a threshold. As their distance becomes smaller, the penalty linearly grows until a terminal threshold is reached (Eq. (\ref{eq:rewards})). Similarly, for the \emph{Boundary} rule, the penalty is calculated based on the distance between the intelligent cell and the eggshell. Finally, for the \emph{Destination} rule, bigger positive rewards are given as the cell moves towards the destination. Details are demonstrated in Supplementary Material S1.2.

\begin{equation}
r = \frac{d-d_l}{d_h-d_l} \times (r_h-r_l) + r_l,
\label{eq:rewards}
\end{equation}
where d is the distance between two cells and $d_h$ and $d_l$ represent the highest and lowest bounds of the distance between two cells where a penalty is generated. $r_h$ and $r_l$ indicate the range of the penalty.

\subsection{An Agent-based Deep Reinforcement Learning Framework for \emph{C. elegans} Embryogenesis}

\begin{figure}[!tpb]
\centering
\subfigure[]{
\includegraphics[width=0.15\textwidth]{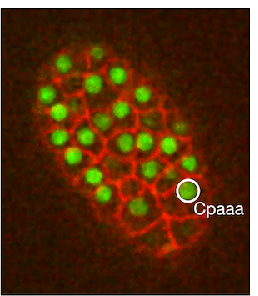}
\includegraphics[width=0.15\textwidth]{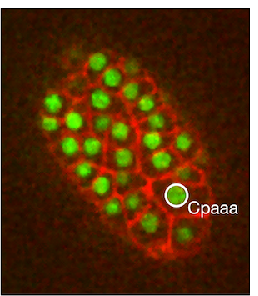}
\includegraphics[width=0.15\textwidth]{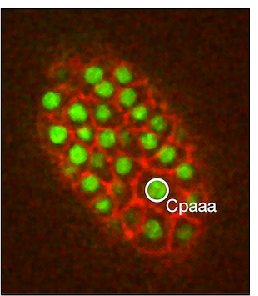}
}
\subfigure[]{
\includegraphics[width=0.15\textwidth]{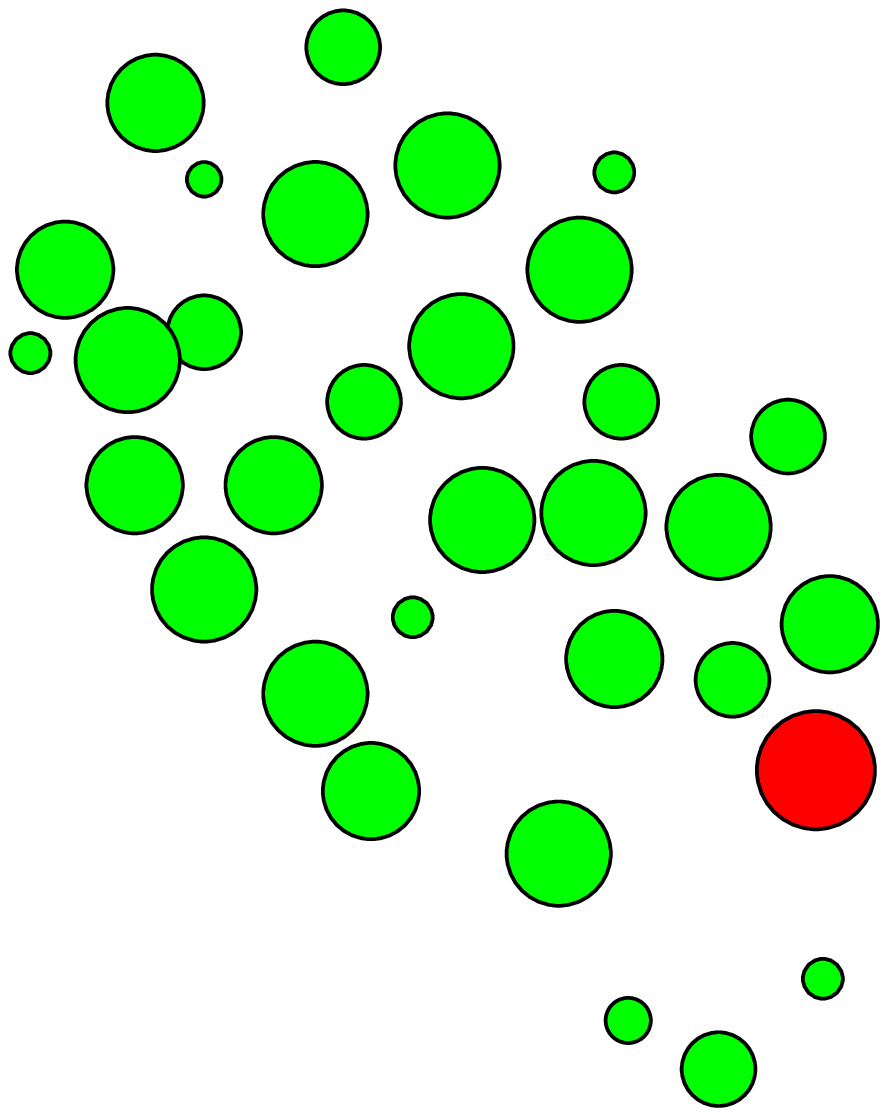}
\includegraphics[width=0.15\textwidth]{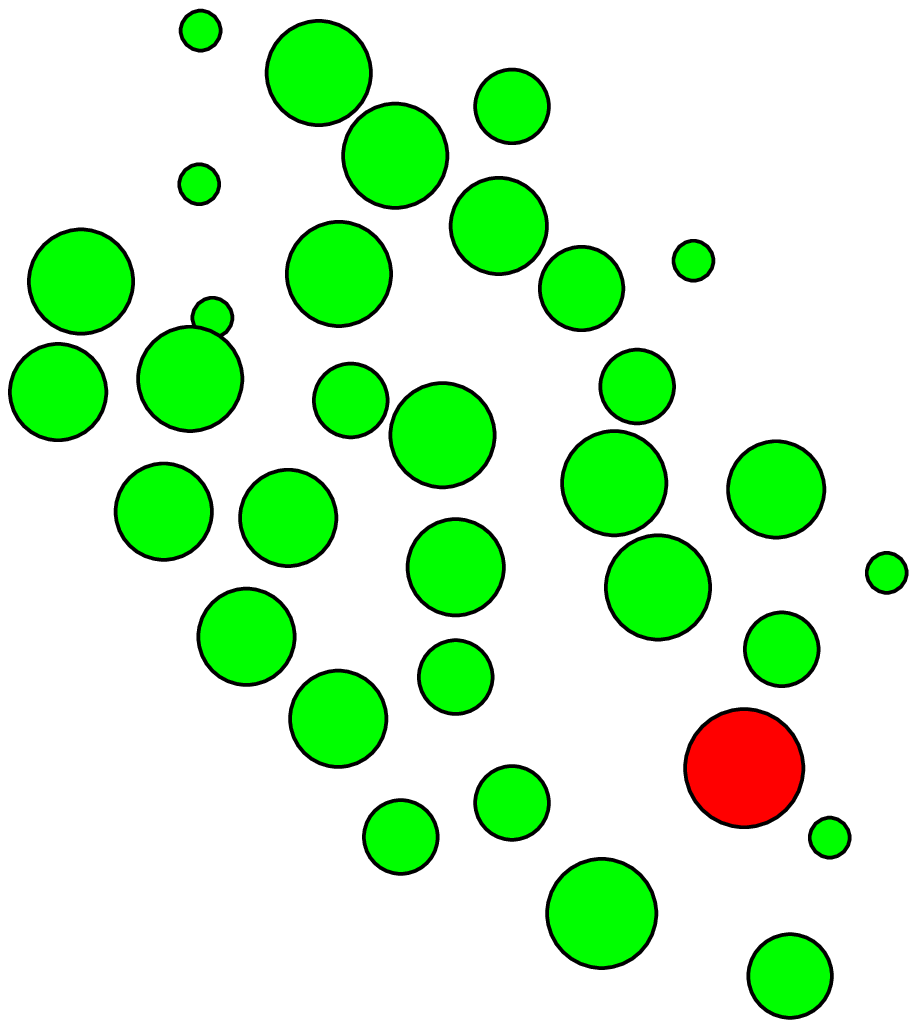}
\includegraphics[width=0.15\textwidth]{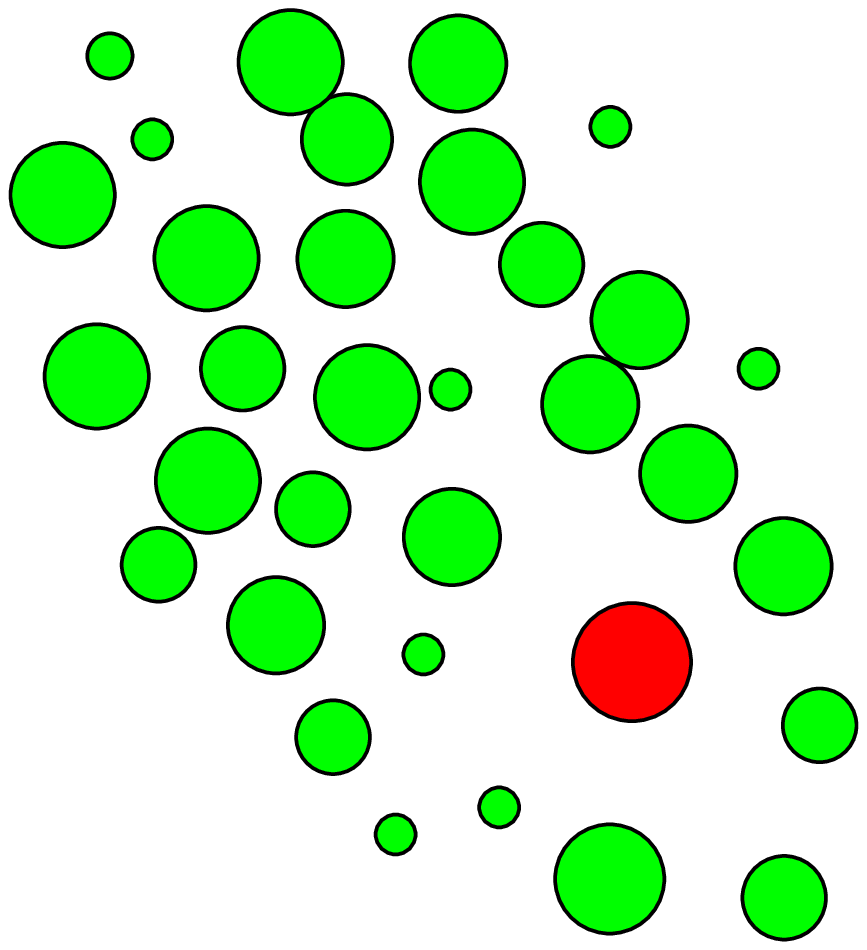}
}
\caption{Comparison between (a) the 3D time-lapse images and (b) the visualizations of the ABM simulation results. Simulation results highly reproduce the observed patterns.}
\label{fig:abm_result}
\end{figure}

The ABM environment was initialized with the observation data from live imaging with automated cell lineage tracing. We first tested the performance of our ABM framework. The ABM platform was configured to track the movements of the intercalation cell, namely, Cpaaa in the first process, for the purpose of illustration. Although the embryo we measured had a length of 30 $\mu$m in the dorsal-ventral axis, we only considered the space that is 5-9 $\mu$m to the dorsal side, where Cpaaa's intercalation happens. The entire space was visualized by projecting all cells in this space to the center plane (7 $\mu$m to the dorsal side). Based on the result (Fig. \ref{fig:abm_result}) we found that the movement path of Cpaaa is consistent with that in the 3D time-lapse images. The visualized cell sizes are largely consistent with the observation data, except the fact that a few of them, especially located in the planes that are far away from the center plane, have slightly different sizes visually. However, those differences have an insignificant impact on cell movement modeling. 

\begin{figure*}[!tpb]
\centering
\subfigure[]{
\label{fig:rewards}
\includegraphics[width=0.31\textwidth]{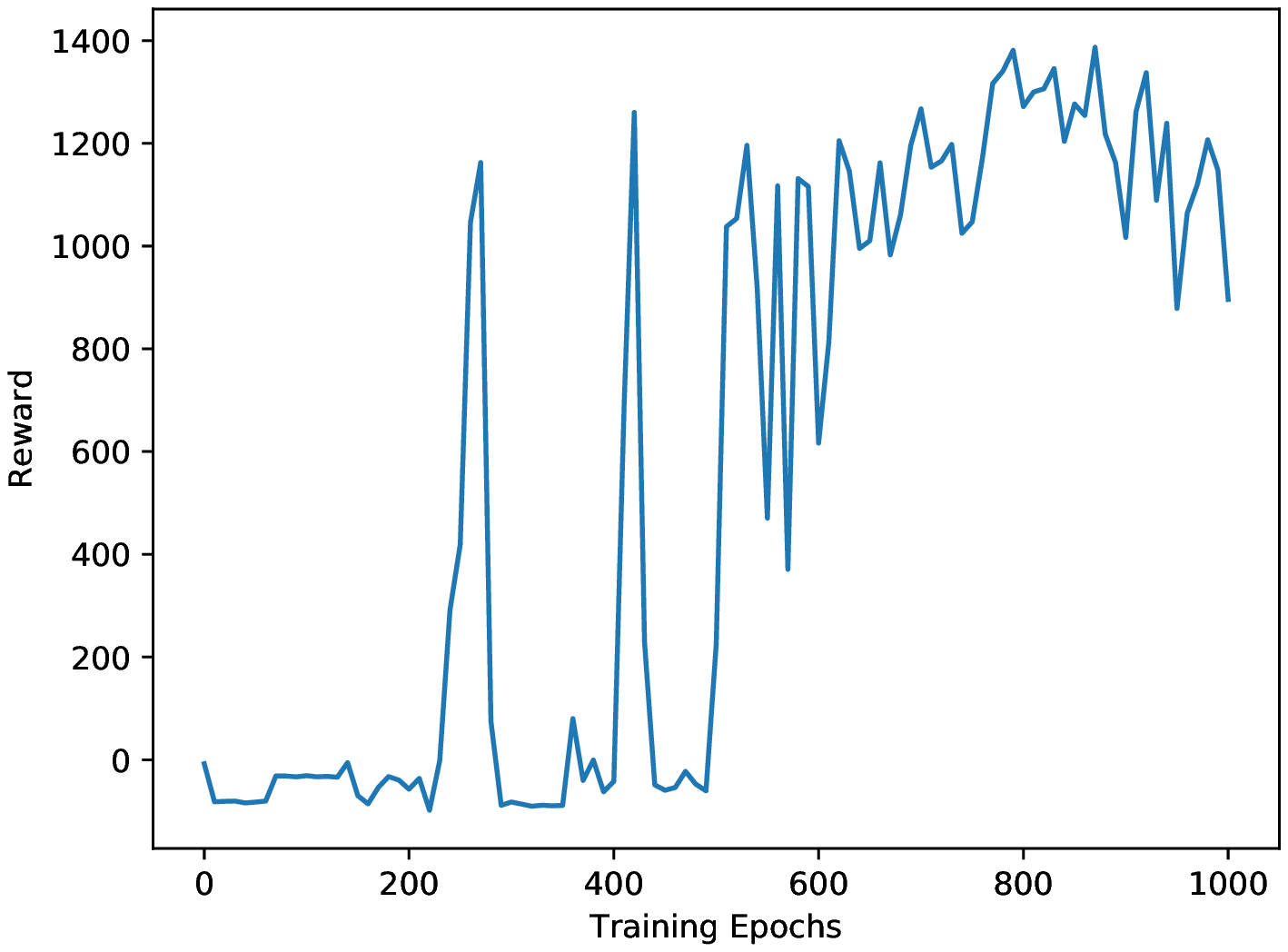}}
\subfigure[]{
\label{fig:loss}
\includegraphics[width=0.31\textwidth]{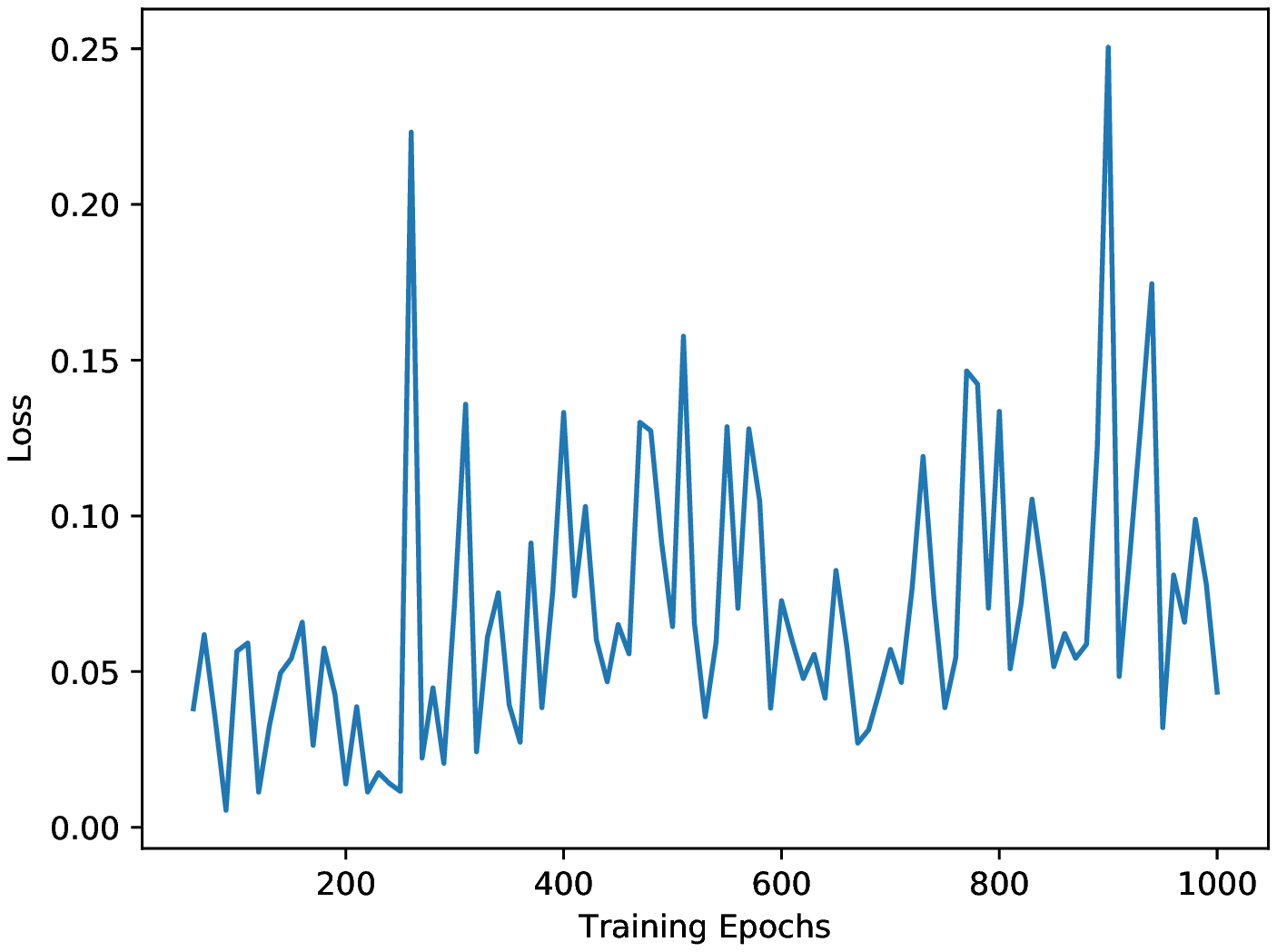}}
\subfigure[]{
\label{fig:action_value}
\includegraphics[width=0.31\textwidth]{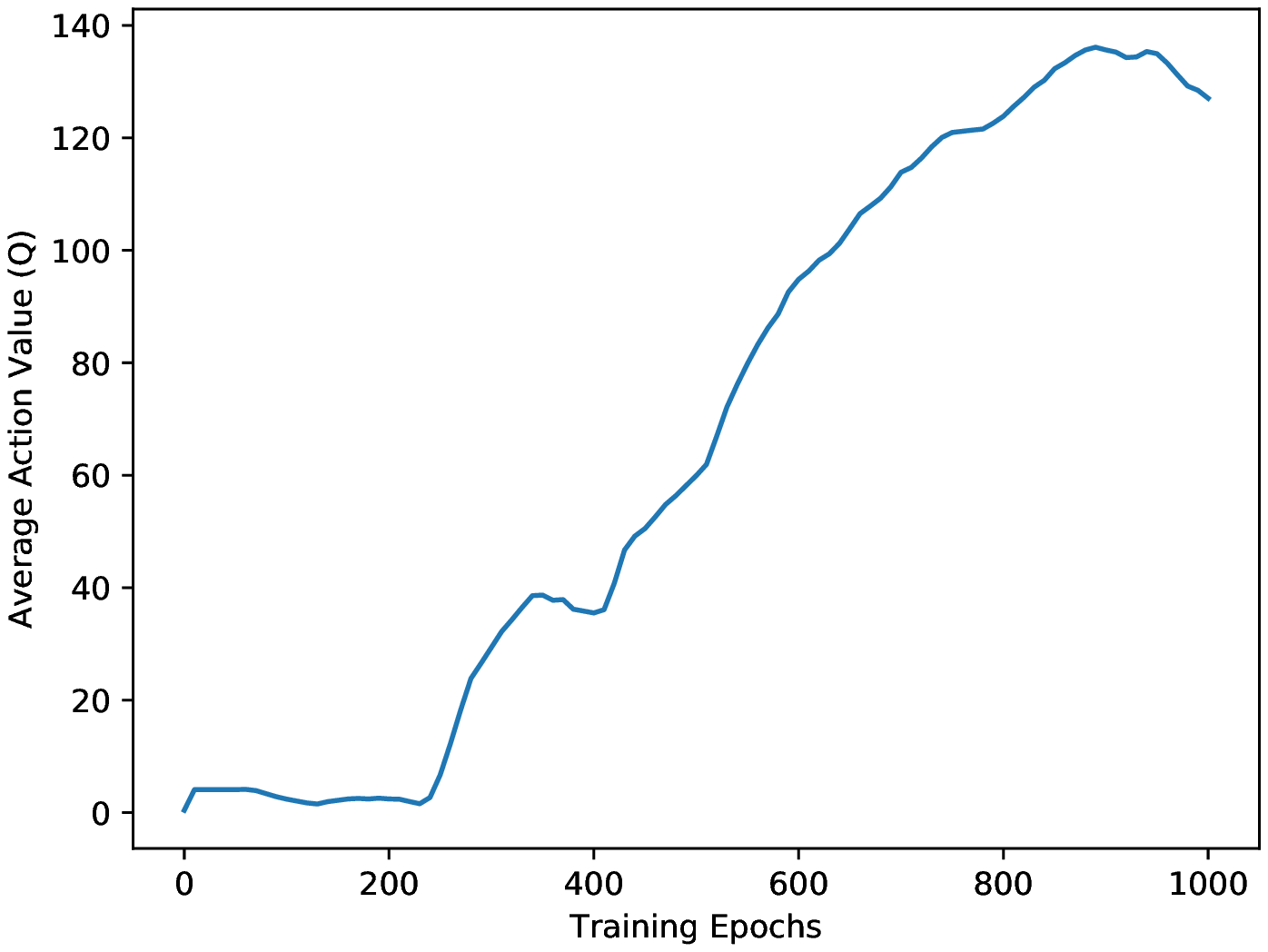}}
\caption{Performance evaluation of the deep reinforcement learning algorithm for cell movement modeling. (a) The accumulated rewards generally goes upward, but tends to be noisy. (b) The loss tends to oscillate because of the implementation of the experience replay and the target network. (c) The average action value grows smoothly over time.}
\label{fig:training}
\end{figure*}

Unlike supervised learning tasks, such as classification and regression, evaluating the performance is quite challenging in deep reinforcement learning tasks. We followed the evaluation metric in \cite{mnih2013playing} to quantify the general performance of the system. The total rewards a cell collects in a single movement path generally goes upward, but tends to be quite noisy since very tiny changes in the weights of the neural network results in large changes in the actions a cell chooses \cite{mnih2013playing} (Fig. \ref{fig:rewards}). Training loss tends to oscillate over time (Fig. \ref{fig:loss}), and the reason behind this is the implementation of the experience replay and the target network, which cut off the correlation between training samples. Finally, we extracted a set of states by running the model in a non-reinforcement learning way and collecting the state cells' locations. We then fed these predefined states to the neural network during the training process. It turns out that the average action values of these states grows smoothly during the training process (Fig. \ref{fig:action_value}). We did not encounter any divergence problems, though the convergence of DQN is still an active research area. Sometimes, we experienced a few unstable training scenarios, but these problems could be solved by implementing a learning rate decay strategy.

\subsection{Regulatory Mechanisms of Individual Cell Movements}
We examined our hypotheses of individual cell movement in the {\em Cpaaa intercalation} case (see Section \ref{sec:individual}). Specifically, we tested (1) whether Cpaaa's intercalation results from an active directional movement or a passive movement, and (2) whether a passive movement mechanism is sufficient for explaining the migration path of Cpaaa's neighbors. In this case, the observed fact is that during the first four minutes of the process, the intercalating cell Cpaaa moves randomly. After extensive divisions of the ABarp cells, Cpaaa changes its behavior to a directional movement until the end of the process. The signal triggering the switch may come from the newborn ABarp cells. 

In the directional cell movement process, unexpected regularization signals or irregular movement patterns have to be considered. In our study, we defined the possibility of selecting a directional movement from the neural network by a ratio between 0 and 1. The value of zero means a completely random movement, and the value of one means a completely directional cell movement.

\subsubsection{Regulatory mechanisms in the {\em Cpaaa intercalation} case}
\begin{figure*}[!tpb]
\centering

\begin{minipage}{2.1\columnwidth}
\centering
\subfigure[]{
\includegraphics[width=0.15\textwidth]{168_sf.eps}
\includegraphics[width=0.15\textwidth]{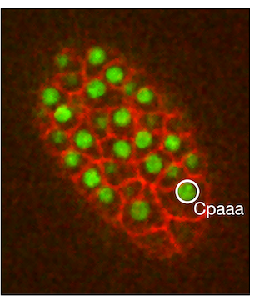}
\includegraphics[width=0.15\textwidth]{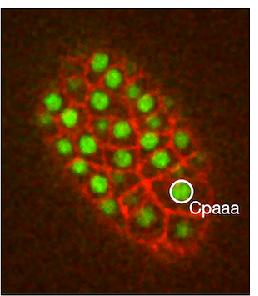}
\includegraphics[width=0.15\textwidth]{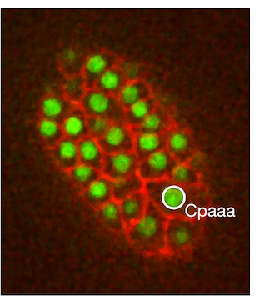}
\includegraphics[width=0.15\textwidth]{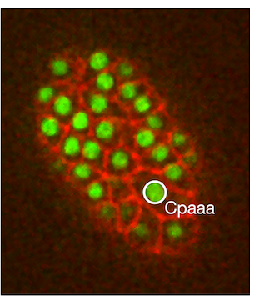}
\includegraphics[width=0.15\textwidth]{190_sf.eps}}
\end{minipage}

\begin{minipage}{2.1\columnwidth}
\centering
\subfigure[]{
\includegraphics[width=0.15\textwidth]{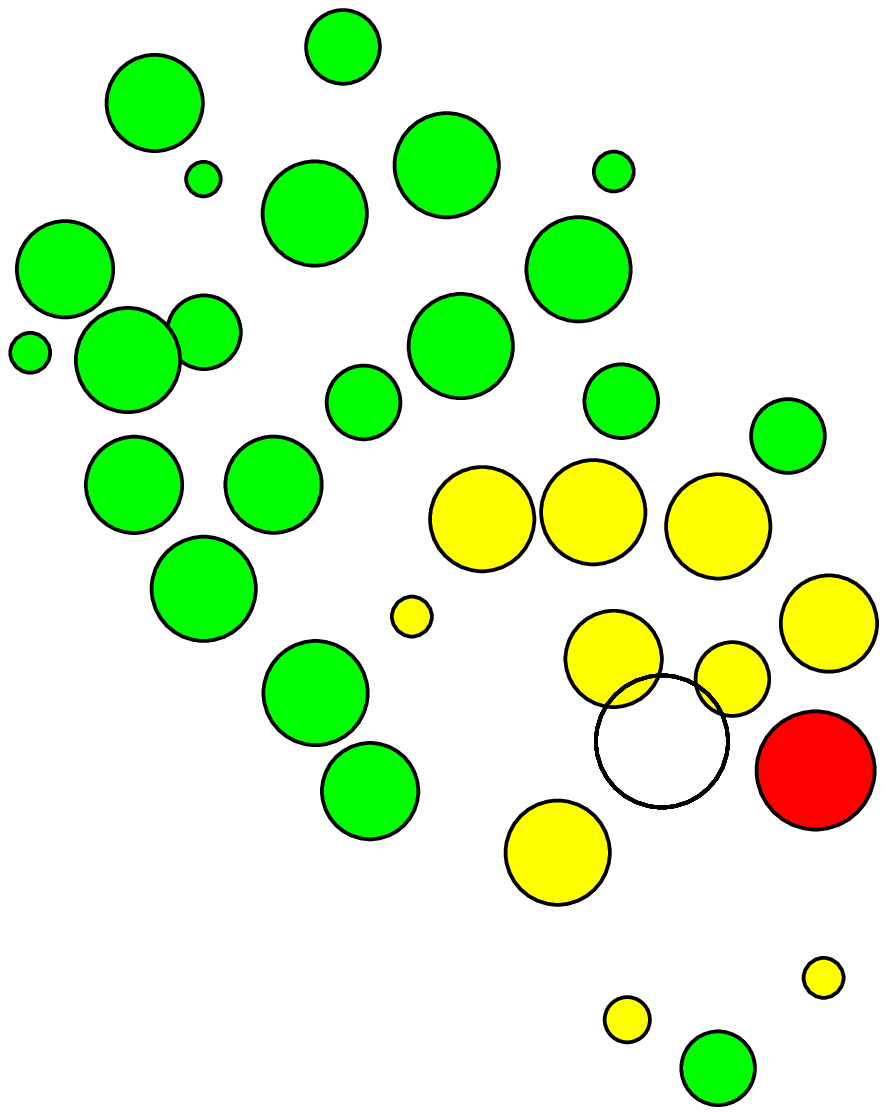}
\includegraphics[width=0.15\textwidth]{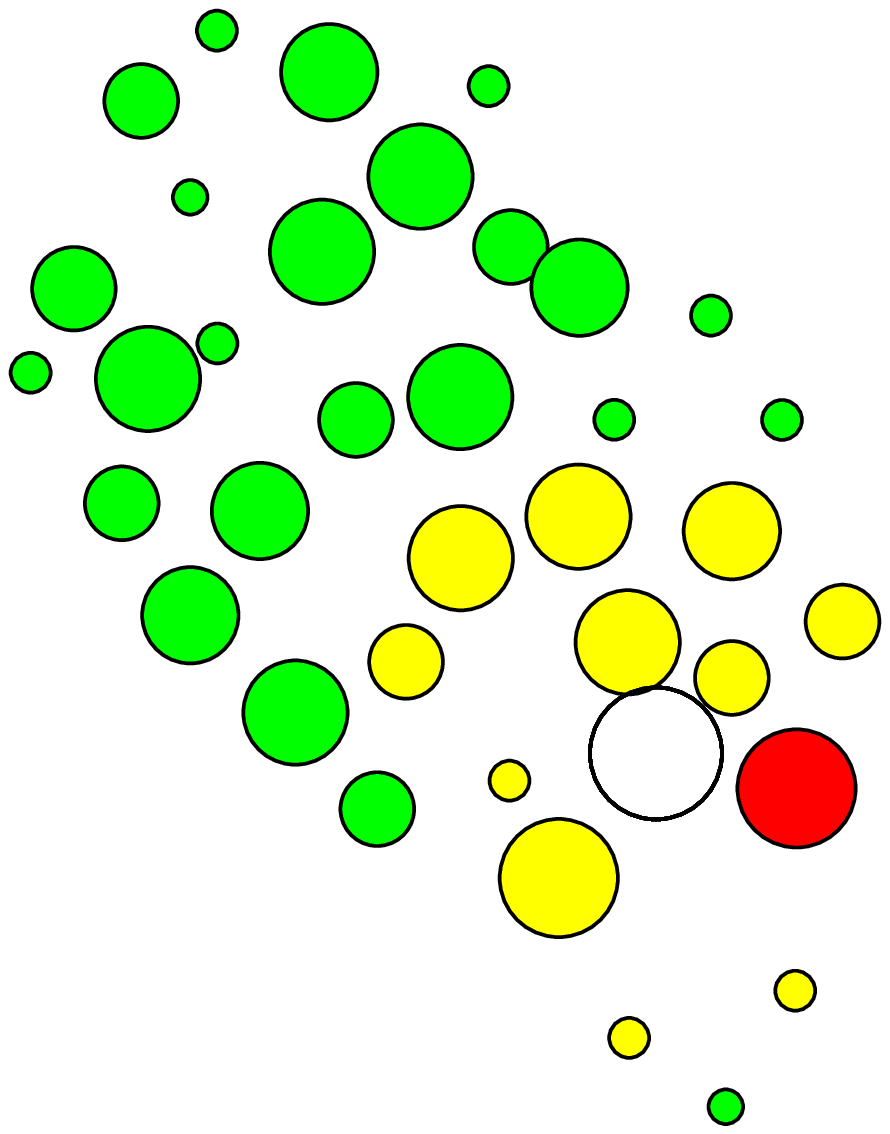}
\includegraphics[width=0.15\textwidth]{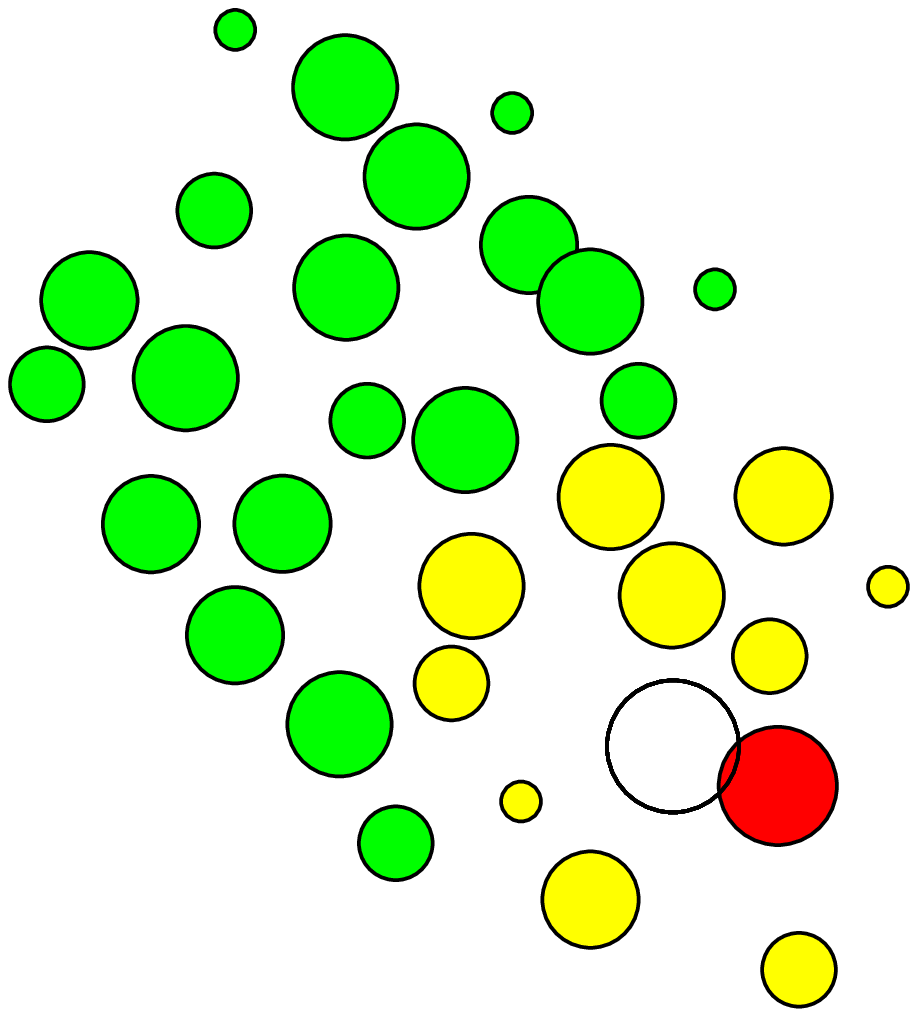}
\includegraphics[width=0.15\textwidth]{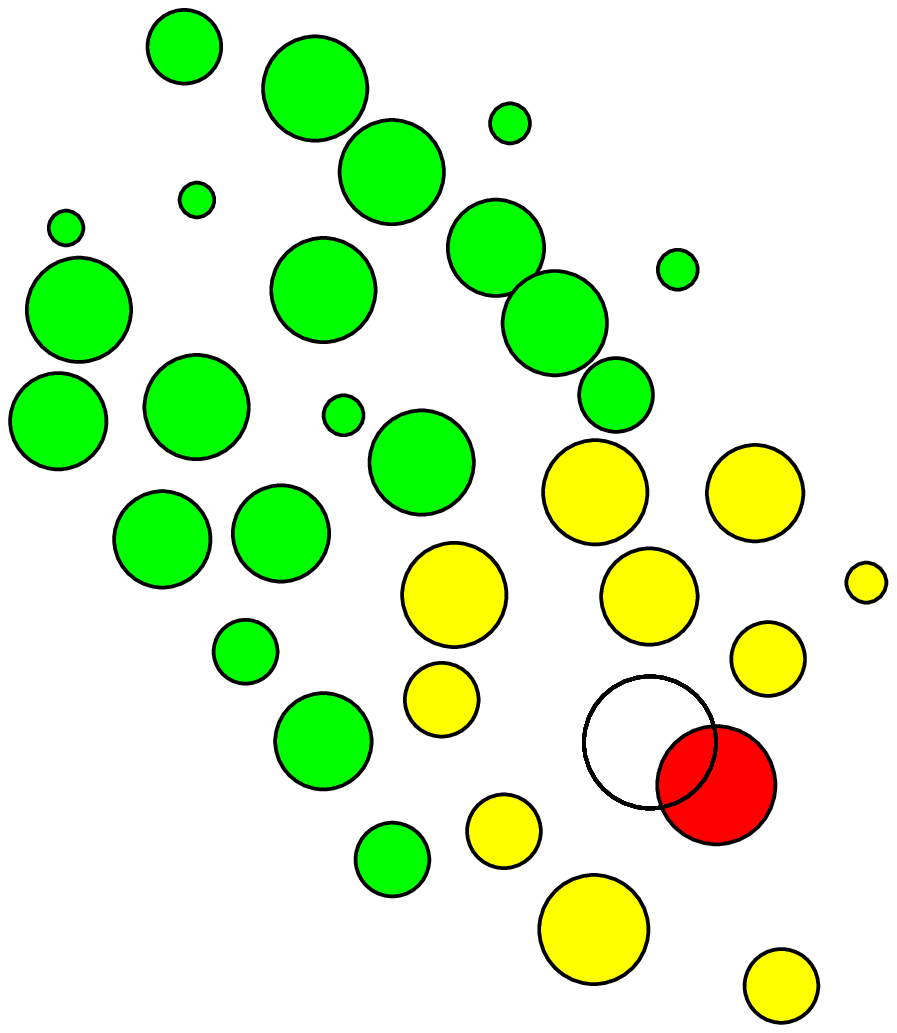}
\includegraphics[width=0.15\textwidth]{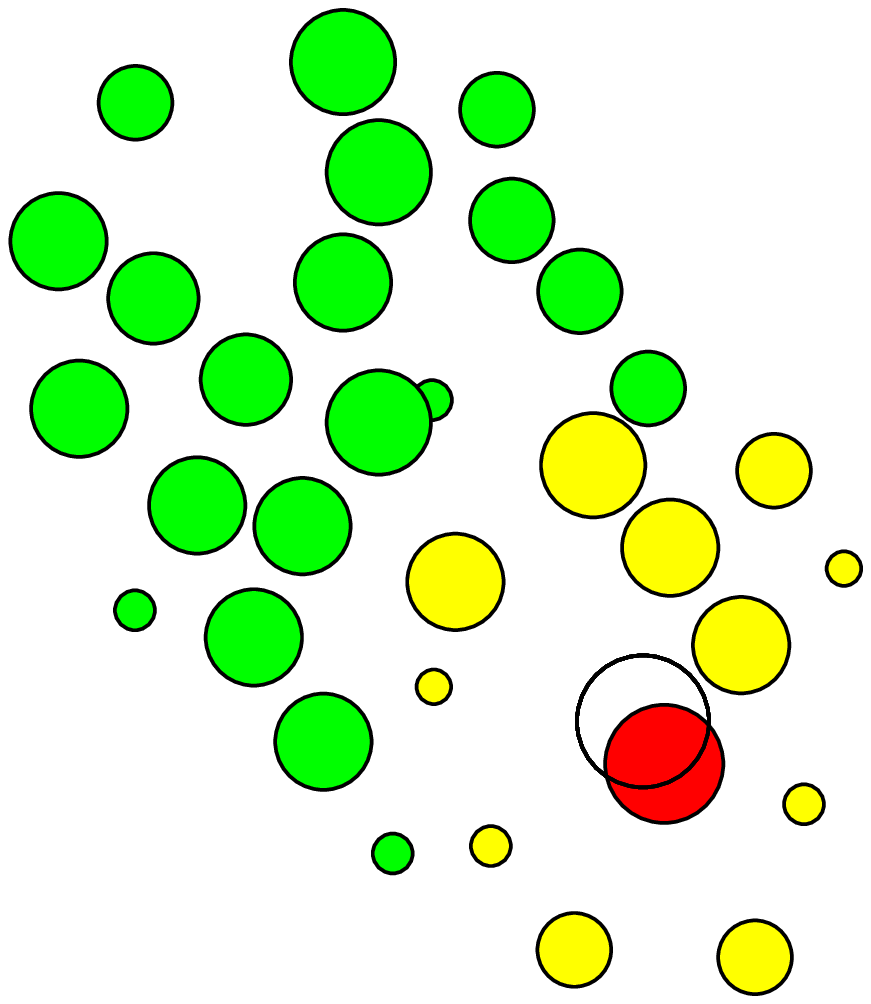}
\includegraphics[width=0.15\textwidth]{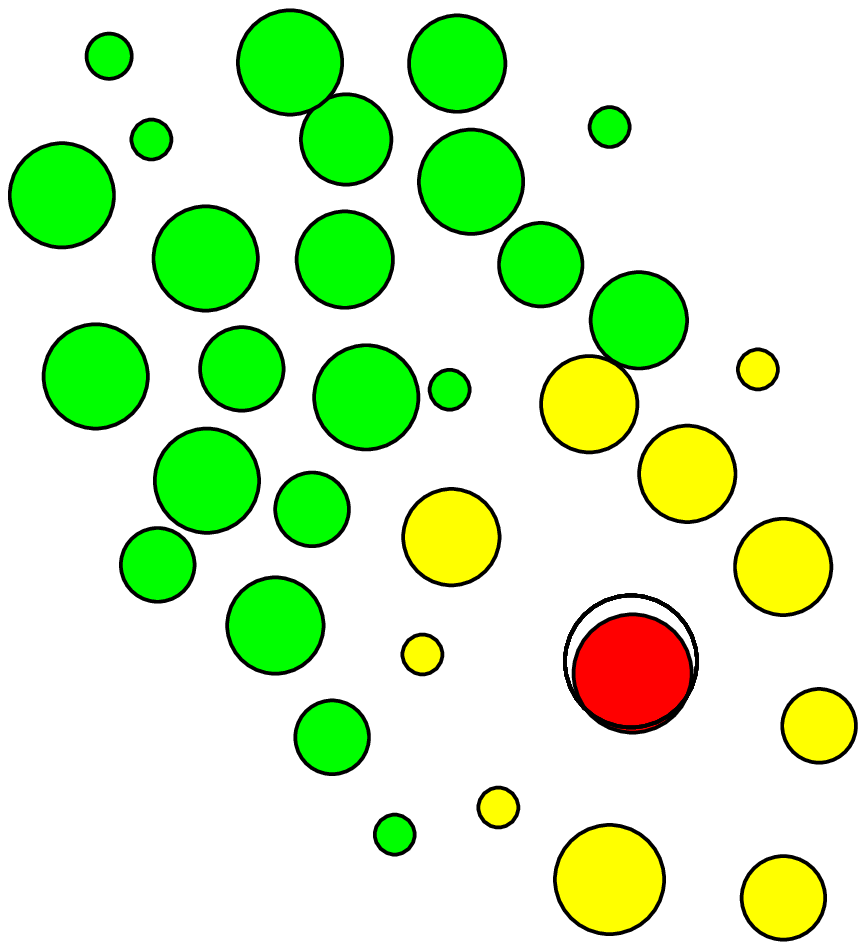}
}
\end{minipage}

\begin{minipage}{2.1\columnwidth}
\centering
\subfigure[]{
\includegraphics[width=0.15\textwidth]{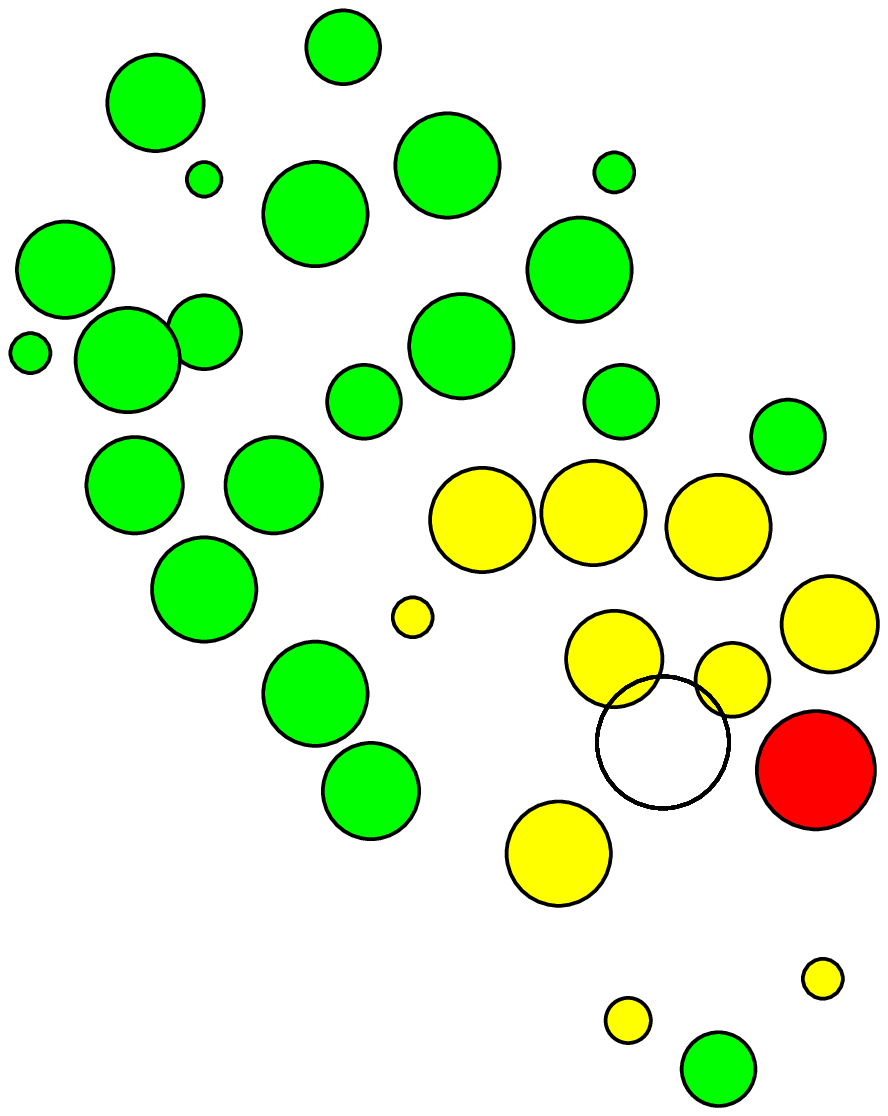}
\includegraphics[width=0.15\textwidth]{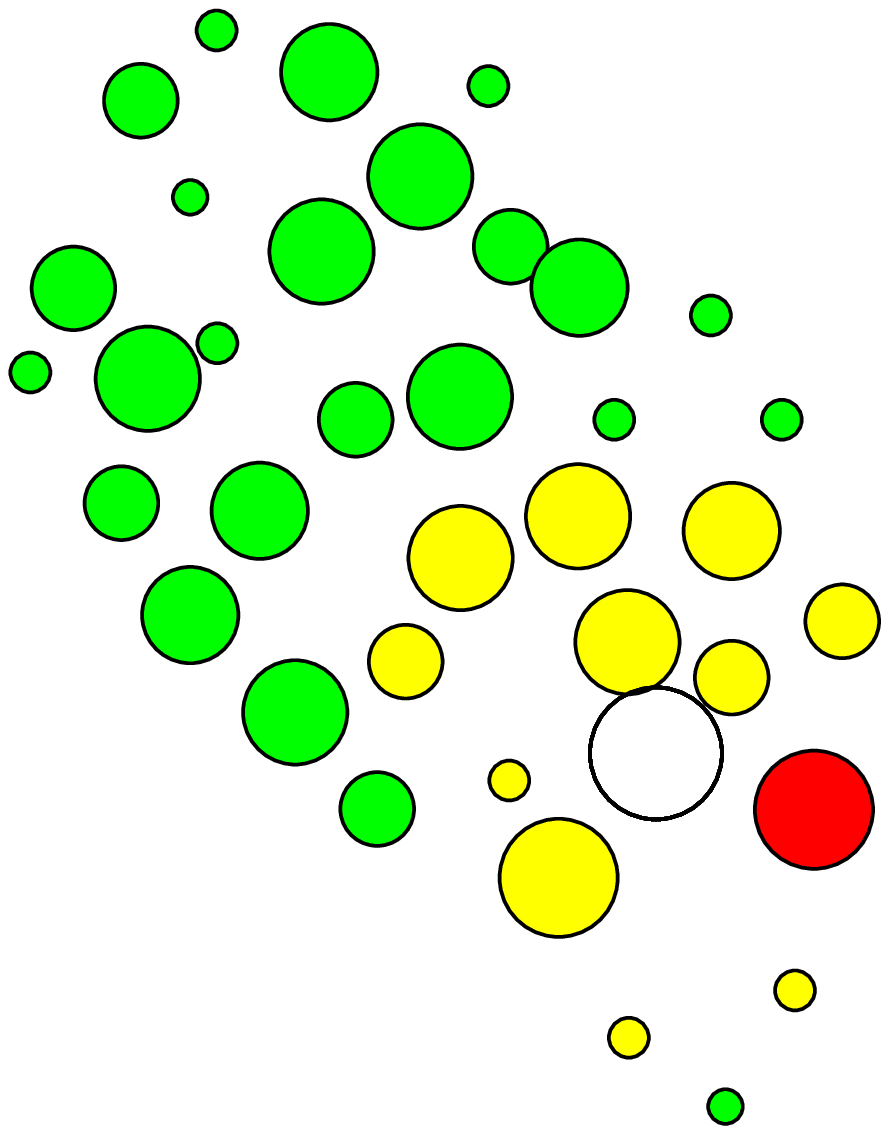}
\includegraphics[width=0.15\textwidth]{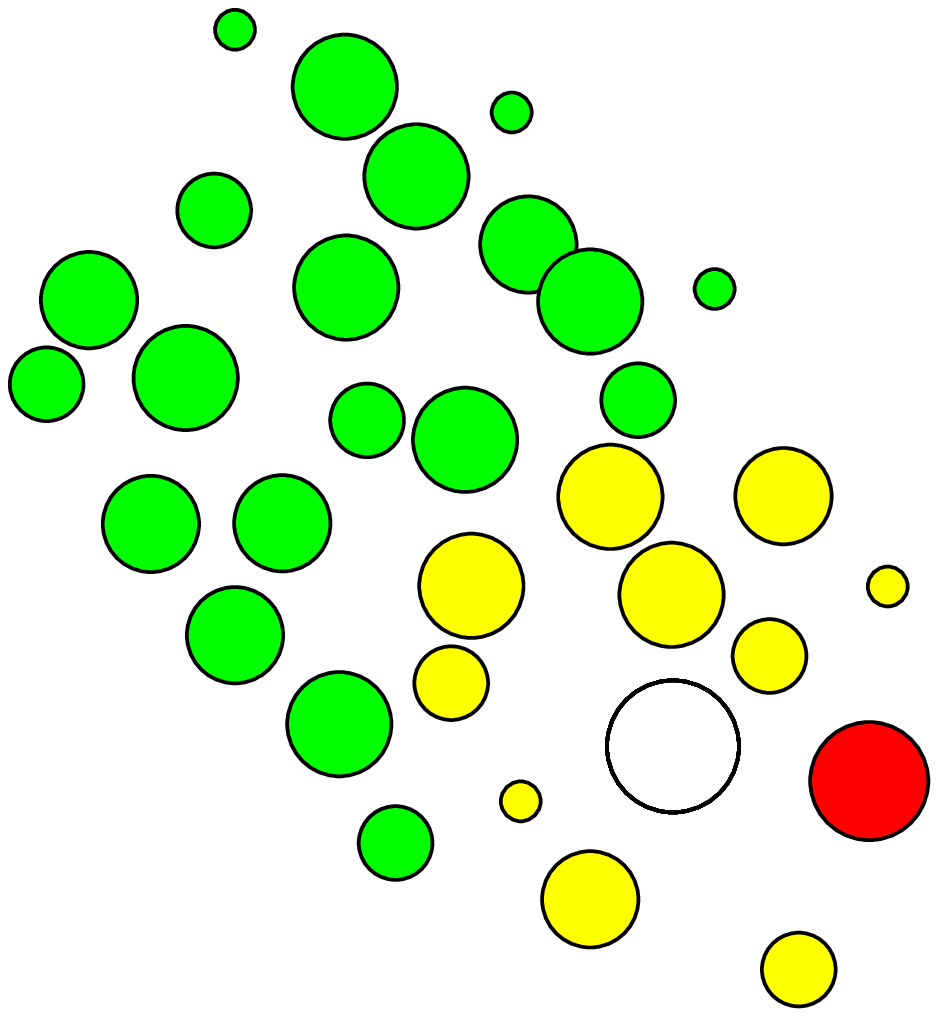}
\includegraphics[width=0.15\textwidth]{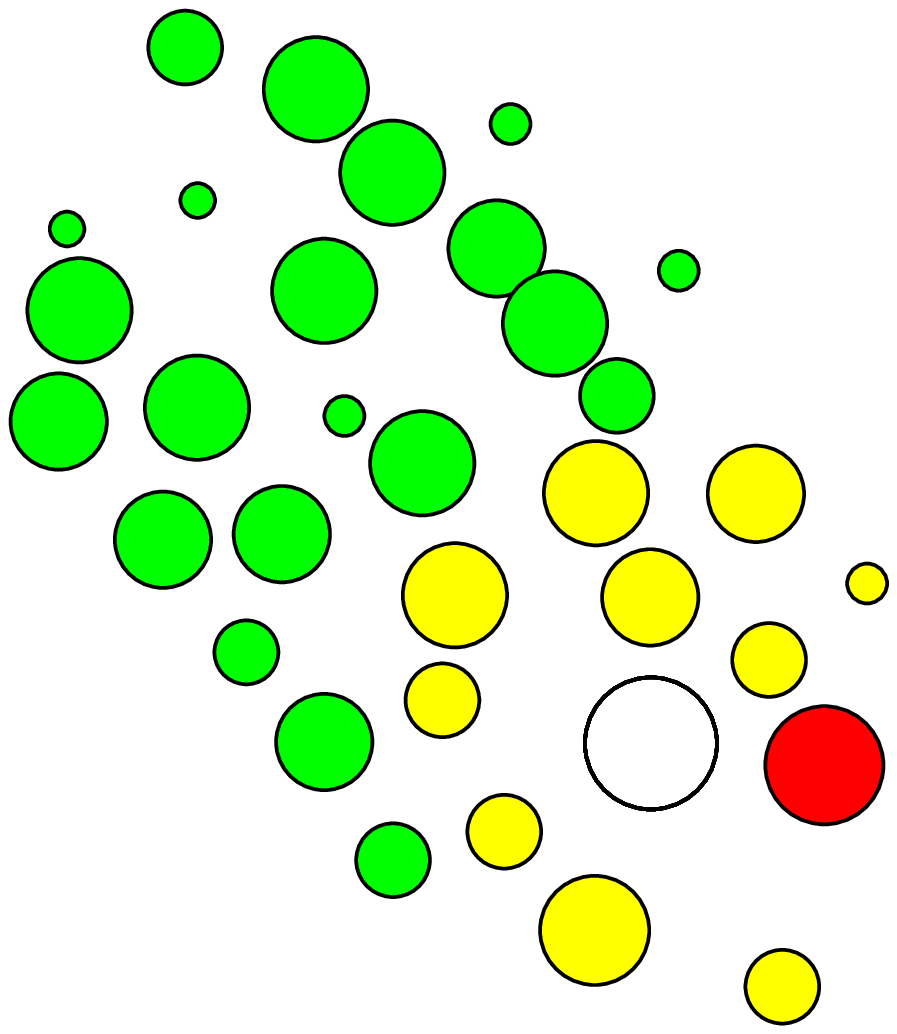}
\includegraphics[width=0.15\textwidth]{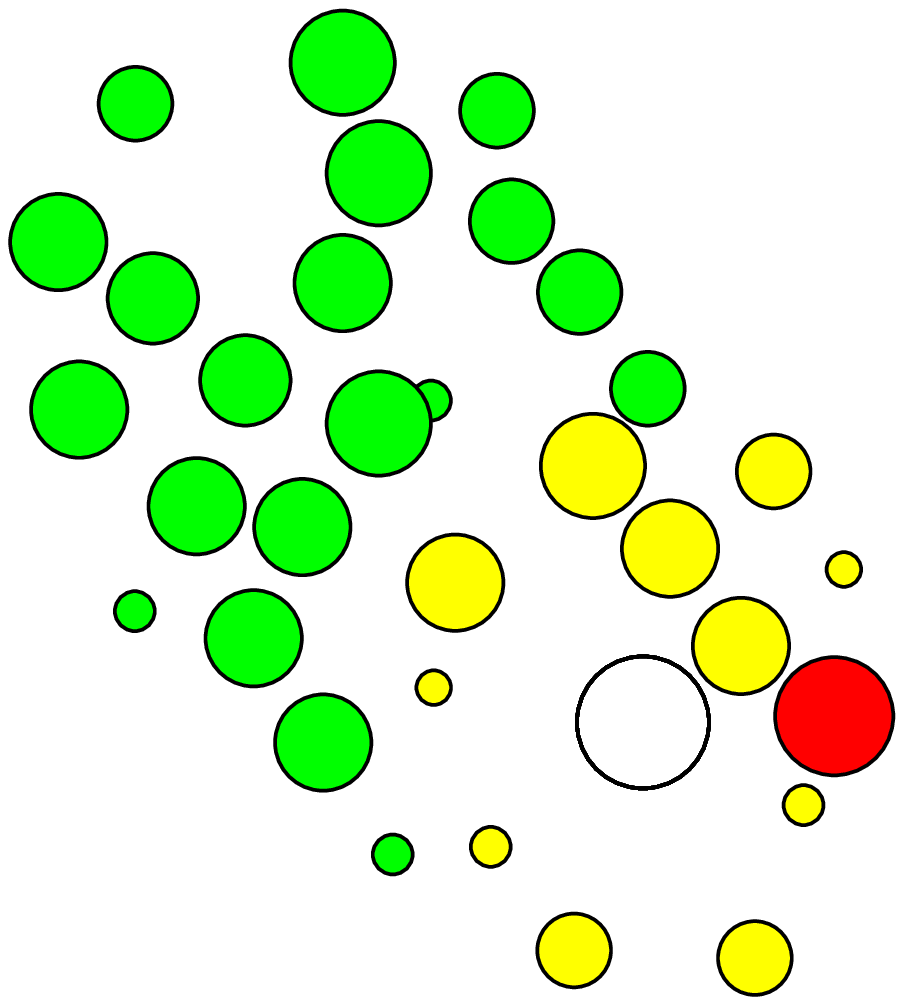}
\includegraphics[width=0.15\textwidth]{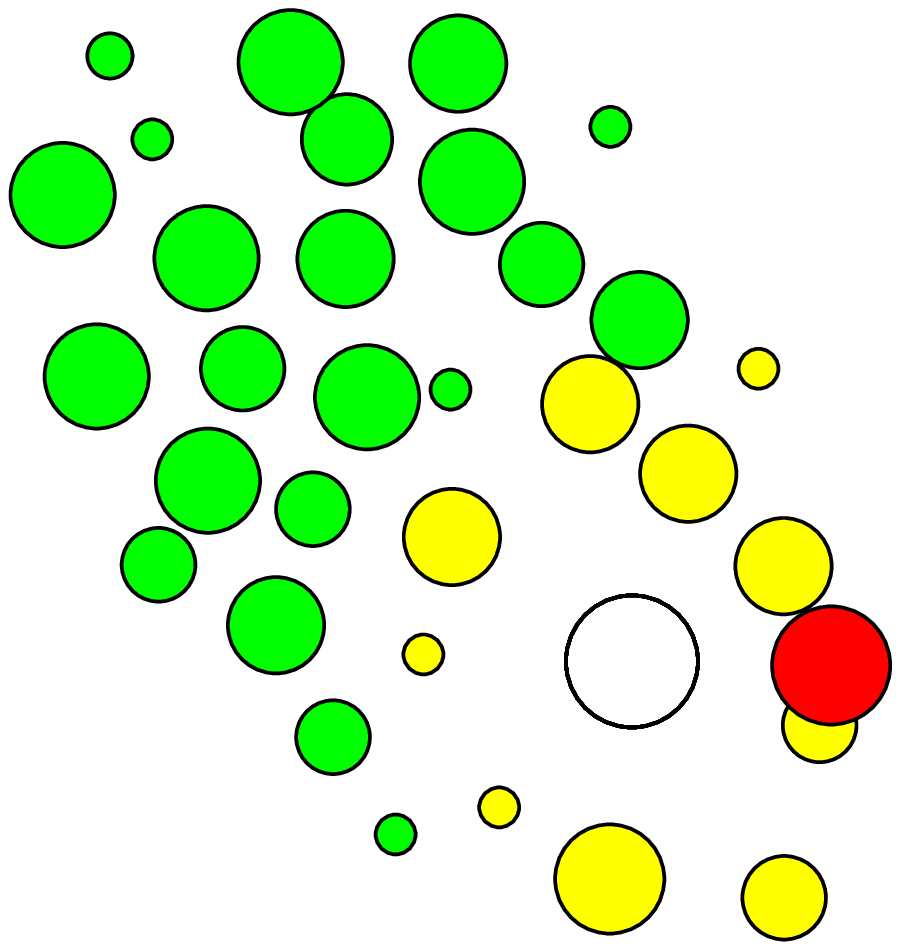}
}
\end{minipage}

\begin{minipage}{2.1\columnwidth}
\centering
\subfigure[]{
\includegraphics[width=0.15\textwidth]{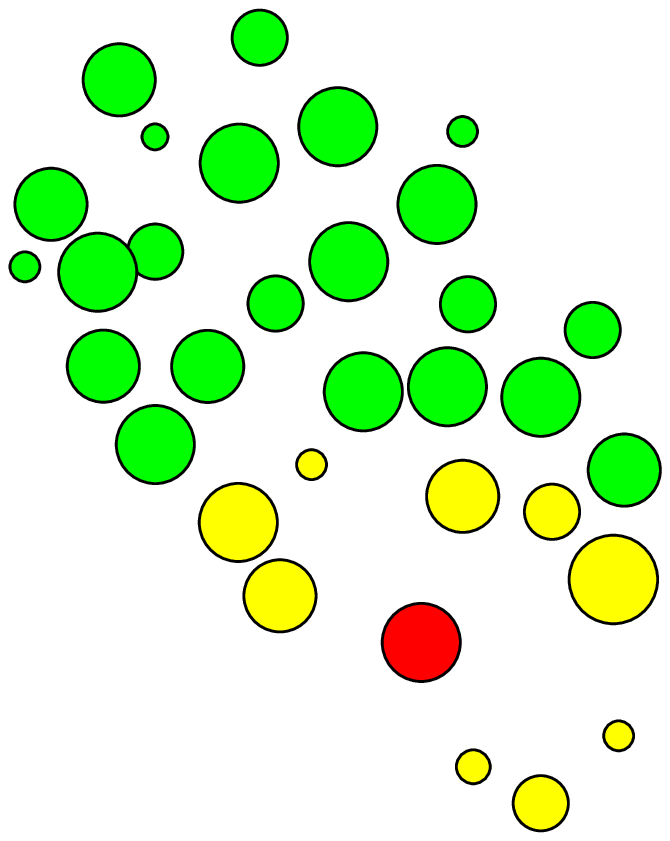}
\includegraphics[width=0.15\textwidth]{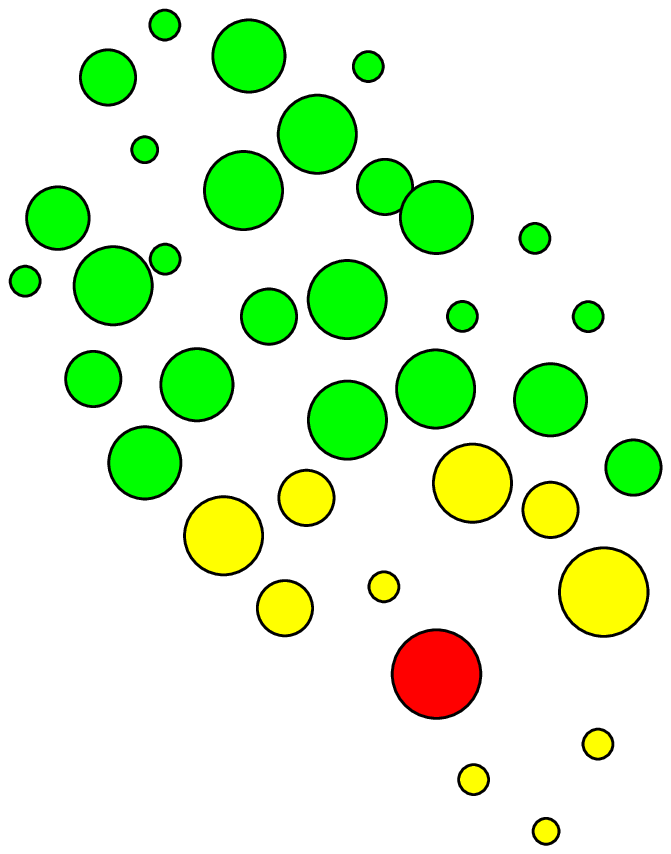}
\includegraphics[width=0.15\textwidth]{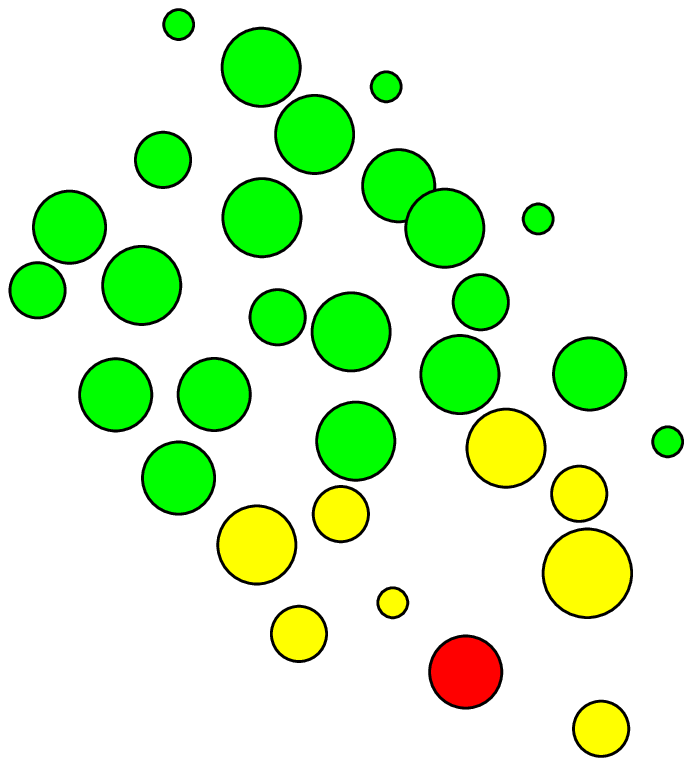}
\includegraphics[width=0.15\textwidth]{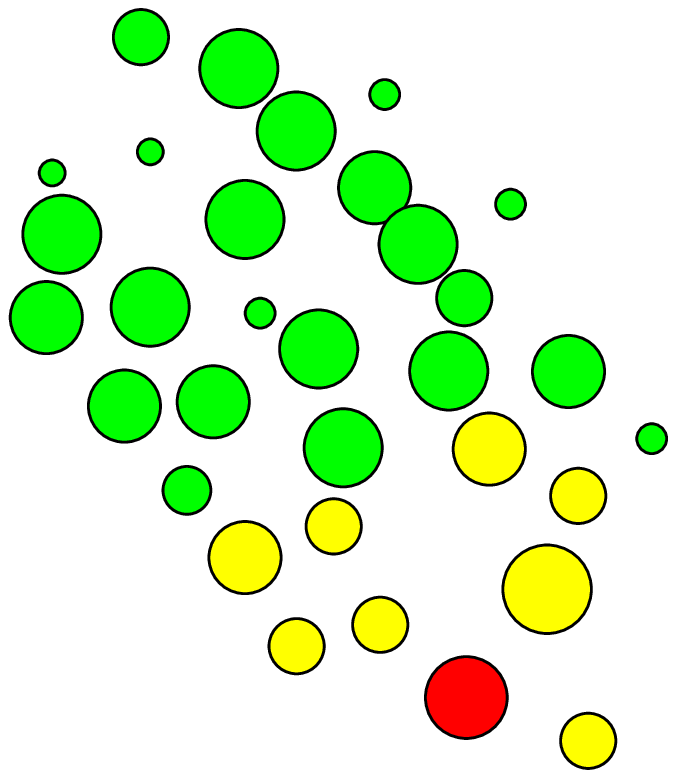}
\includegraphics[width=0.15\textwidth]{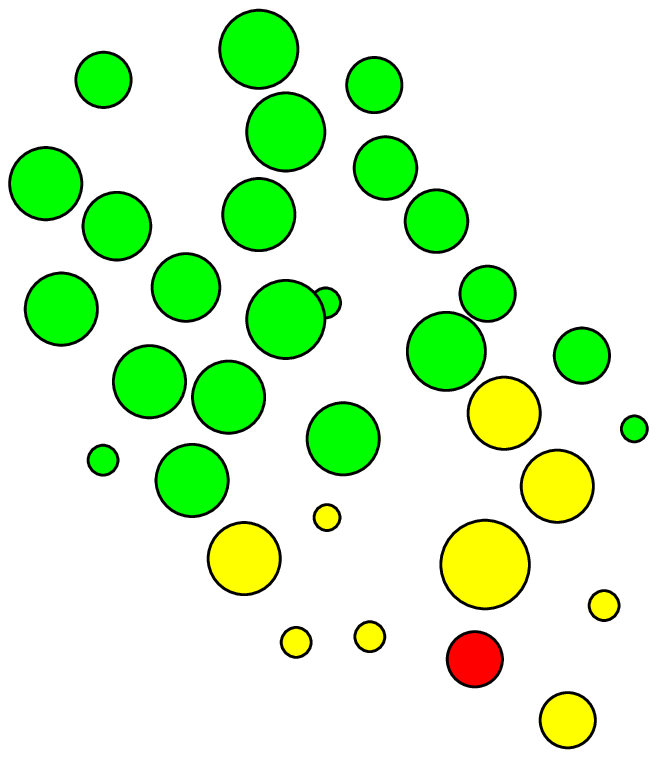}
\includegraphics[width=0.15\textwidth]{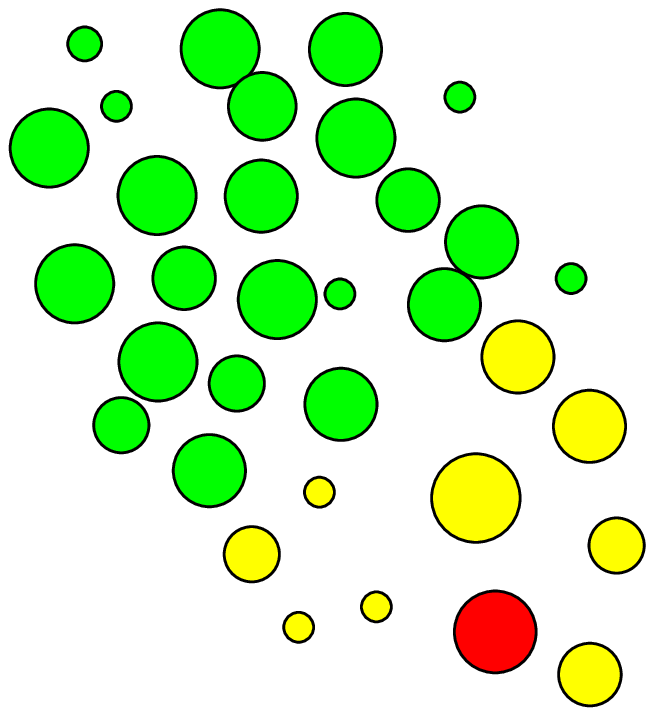}
}
\end{minipage}
\caption{Results of the {\em Cpaaa intercalation} case. (a) Observation results visualized by Acetree from 3D time-lapse images. (b) Simulation results of the intercalating cell Cpaaa with the {\em Destination} rule. (c) Simulation results when training Cpaaa only with the {\em Boundary} and {\em Collision} rules, without the {\em Destination} rule, which indicate that Cpaaa fell into a suboptimal location. (d) Simulation results of the cell Caaaa, a neighbor of Cpaaa. Red, yellow, and green circles represent the intelligent cell, input state cells, and non-related cells, respectively. The white circle indicates the destination of the intelligent cell. All four sets of data were collected at the following time steps: 0, 4, 8, 12, 17, and 22 (minutes from the beginning of the simulation).}
\label{fig:result_rosette}
\end{figure*}

We trained individual neural networks (parameters were initialized by random sampling from a Gaussian distribution.) for directional and passive movements with different sets of regulatory mechanisms. Specifically, we trained the neural network for passive movement with the {\em Collision} and {\em Boundary} rules, and the one for directional movement with an addition of the {\em Destination} rule. The different behaviors of Cpaaa (random movement for the first four minutes and directional movement after that) were controlled by manipulating the probability of random movement $\epsilon$ in the action selection procedure. The results of the simulation of Cpaaa with the {\em Destination} rule (Fig. \ref{fig:result_rosette}(b)) show that during the first four minutes, the intelligent cell didn't have an explicit destination and, to a large extent, acted randomly. After that, Cpaaa switched its behavior and began to move directionally to the destination, as well as kept proper distances from its neighbors and the eggshell. The whole migration path largely reproduced that in the live microscopy images (Fig. \ref{fig:result_rosette}(a)). However, when we trained Cpaaa without the {\em Destination} rule, it failed to identify the migration path and fell into a suboptimal location where it kept proper distances with its neighbors (Fig. \ref{fig:result_rosette}(c)). We also trained a neighbor of Cpaaa, namely, Caaaa, as a passive movement cell during the process (Fig. \ref{fig:result_rosette}(d)), and its migration path in this scenario also reproduced that in the images, which indicated that Caaaa played a passive role during Cpaaa's intercalation.

For the verification of the generality of the model, random noises were added to the initial positions of all the cells (including the intelligent cell) and to all the migration paths of the dumb cells during the training process. It turns out that the neural networks could still provide the most proper actions under a large variety of input states after the policy converges, though the optimization process took longer to converge than that in the scenarios without random noises. 

\subsubsection{Migration path of the intelligent cell}
\begin{figure}[!tpb]
\centering
\subfigure[]{
\includegraphics[width=0.22\textwidth]{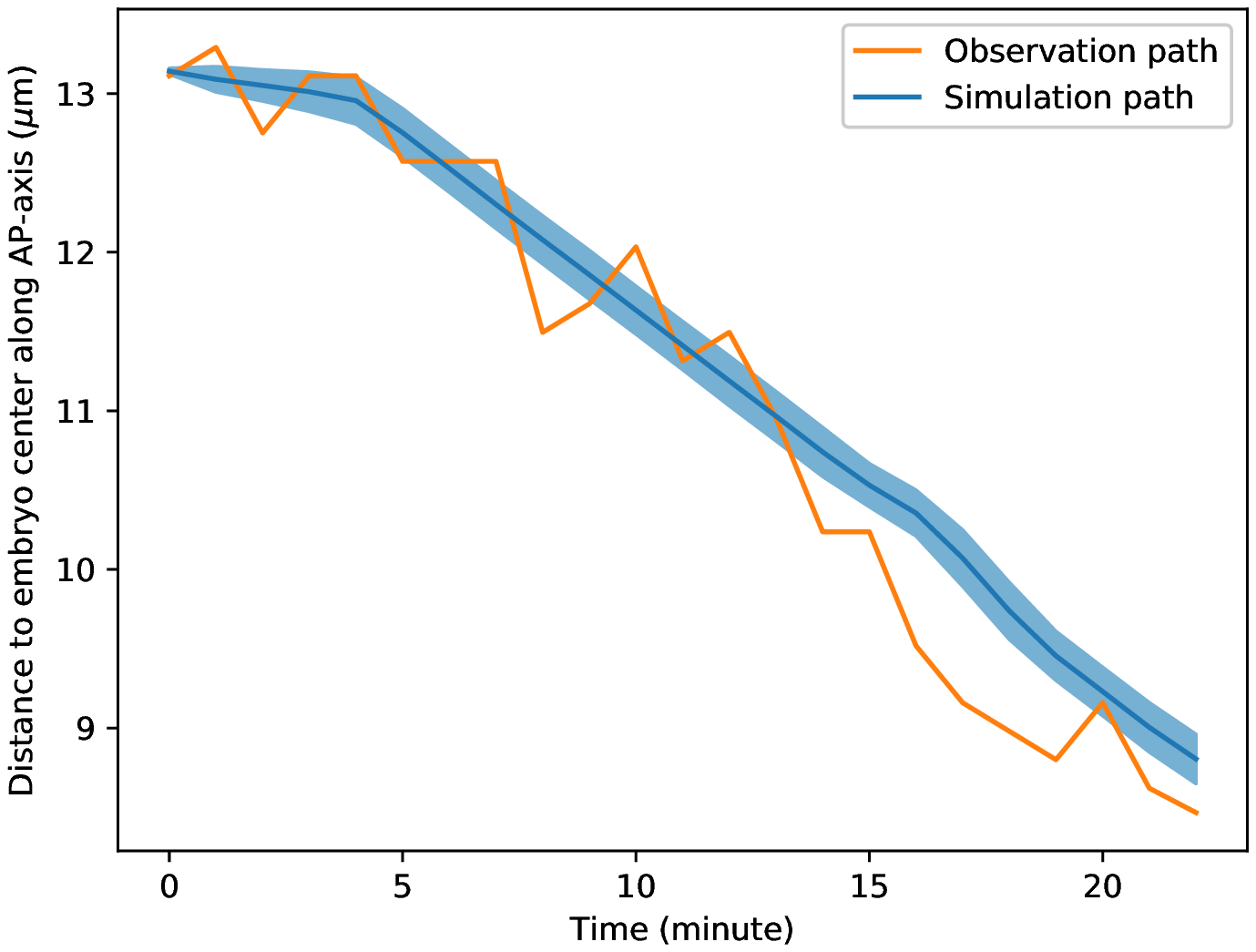}}
\subfigure[]{
\includegraphics[width=0.22\textwidth]{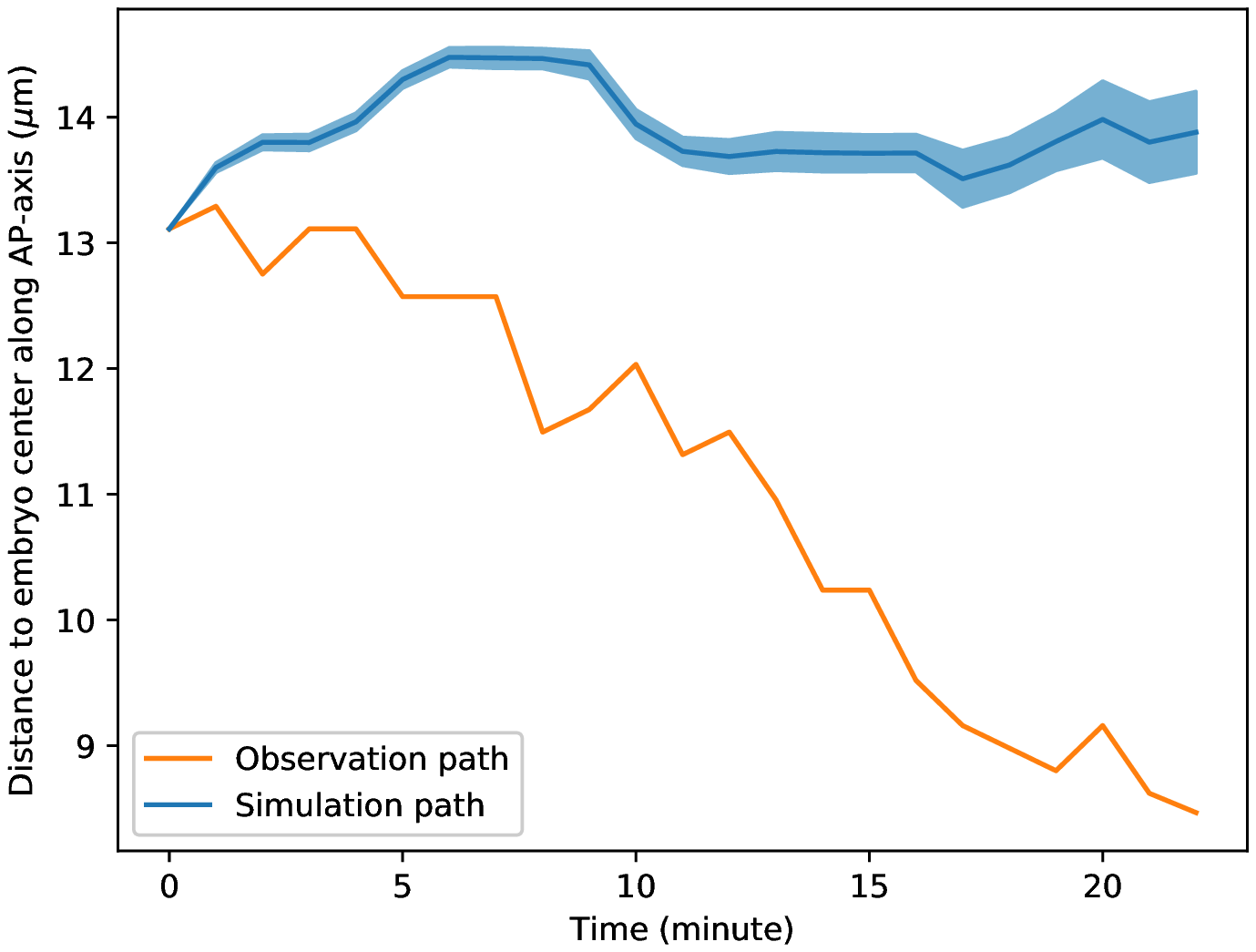}}
\caption{(a) Migration paths of Cpaaa with directional movement. (b) Simulation results when training Cpaaa only with the {\em Boundary} and {\em Collision} rules, without the {\em Destination} rule. Results indicate that Cpaaa fell into a suboptimal location. Both simulation paths are the averages over 50 runs, and the shaded regions indicate ranges of one standard deviation greater/less than the average values. The horizontal axis represents the developmental time in minutes. The vertical axis represents the projected position of Cpaaa on the AP-axis to the center of the embryo. }\label{fig:dist2embryo}
\end{figure}

We found that qualitatively, the intelligent cell Cpaaa adopted a similar migration path to the destination with the directional movement setting, as compared to the observation case (Fig. \ref{fig:dist2embryo}(a)), though from the 13th to 19th minute, the observation movement of Cpaaa went towards the anterior faster than the simulation path. The difference between the simulation and observation results indicates that extra regulatory mechanisms (such as cell adhesion, or intermediate sub-mechanisms, see the Discussion section) could be considered to control cell movement during the whole {\em Cpaaa intercalation} process. On the other hand, without the {\em Destination} rule, Cpaaa's simulated path is quite far away from the observed path (Fig. \ref{fig:dist2embryo}(b)). We used the mean square error (MSE) as a quantitative measurement of the simulated path and the observed path. It turns out that the MSE in Fig. \ref{fig:dist2embryo}(a) is much smaller than that in Fig. \ref{fig:dist2embryo}(b) (4.05 vs. 237.60). In conclusion, the above results show that Cpaaa's intercalation is regulated by an active directional movement mechanism, which is strongly influenced by the {\em Destination} rule (or its alternatives), rather than by a passive movement mechanism. Moreover, another interesting finding is that the standard deviation of the migration path of Cpaaa with the {\em Destination} rule is controlled in a proper range, whereas that of the path without the {\em Destination} rule diverges as time goes by. Such a result indicates that the intelligent cell achieves an error correction mechanism in its migration path to the destination. 

\subsection{Regulatory Mechanisms of Group Cell Migration}
\begin{figure}[!tpb]
\centering
\subfigure[]{
\includegraphics[width=0.11\textwidth]{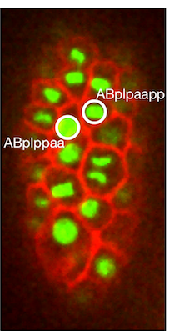}
\includegraphics[width=0.11\textwidth]{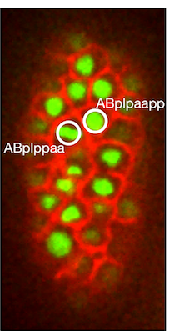}
\includegraphics[width=0.11\textwidth]{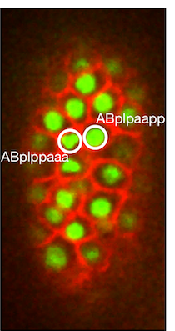}
\includegraphics[width=0.11\textwidth]{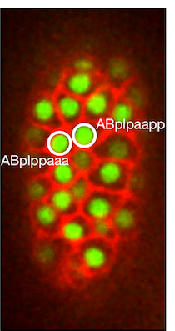}
}
\subfigure[]{
\includegraphics[width=0.11\textwidth]{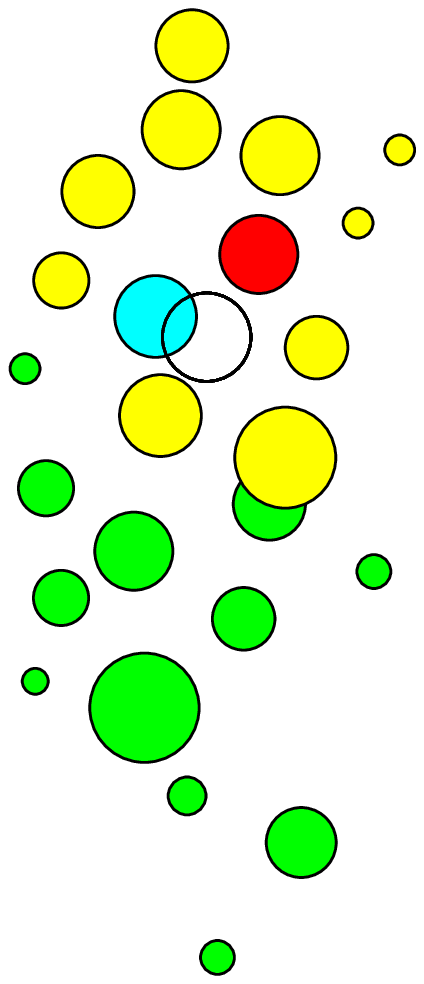}
\includegraphics[width=0.11\textwidth]{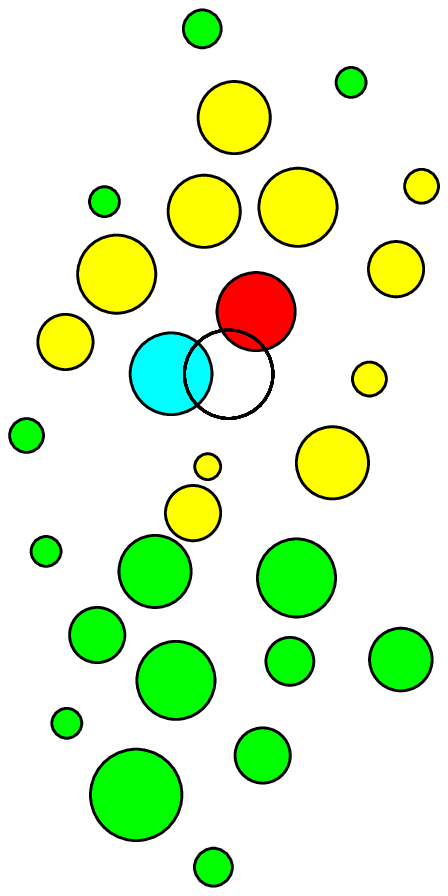}
\includegraphics[width=0.11\textwidth]{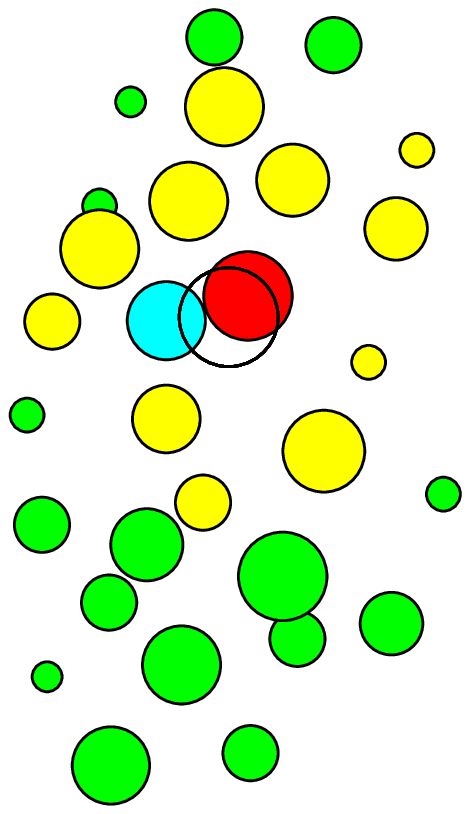}
\includegraphics[width=0.11\textwidth]{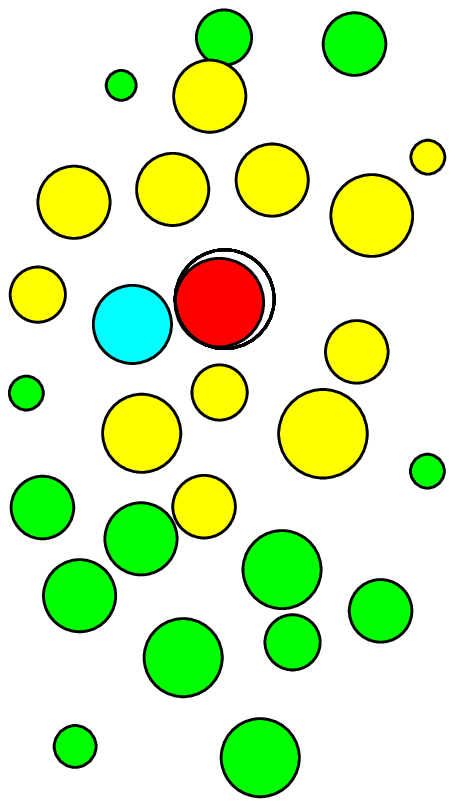}
}
\caption{The simulation of left-right asymmetry rearrangement. (a) Observation data. The intelligent cell and the leading cell are circled. (b) Simulation results. The cyan circle represents the leading cell, and the others are color coded, as in Fig. \ref{fig:result_rosette}. The white circle here indicates the destination of the intelligent cell only for the purpose of visualization. Both sets of data were collected at the following time steps: 0, 3, 6, and 9 (minutes from the beginning of the simulation).}
\label{fig:symres_result}
\end{figure}

In this experiment, we trained the neural network to test the cell movement in group migration via the case of {\em left-right asymmetry rearrangement}. Rather than explicitly pointing out the destination, we let the intelligent cell (ABplpaapp) follow the leading cell (ABplppaa, or its daughter cells). The reward setting was then modified accordingly: When the distance between the leading cell and the following cell is in a proper range, a positive reward is given. The results (Fig. \ref{fig:symres_result}(b)) show that ABplpaapp always moves following the leading cell, and keeps proper distances from its neighbors. Although we did not identify which cell is the leading cell, the intelligent cell will gradually figure out which nearby cell is the leading cell through the training process, because following the leading cell will achieve a big reward. The results are consistent with the observation data (Fig. \ref{fig:symres_result}(a)), which shows the flexibility of our model by replacing the \emph{Destination} rule with more concrete ones.

\section{Discussion}
In this study, we presented a novel approach to model cell movement using deep reinforcement learning within an agent-based modeling framework. Our study showed that neural networks can be adopted to characterize cell movement and that the deep reinforcement learning approach (i.e., DQN) can be used to find the optimal migration path of a cell under certain regulatory mechanisms. As comparing to the heuristic rule-based, agent-based models, with which macroscopical behaviors (such as tissue/organ morphogenesis) can be studied \cite{setty2012multi,setty2008four}, this model provides a new point of view in which single cell movements can be defined and optimized over a considerable period of time. In the \emph{Cpaaa intercalation} case, we tested two hypotheses (active directional movement vs. passive movement) that might explain Cpaaa's migration towards the anterior by manipulating the reward settings (use the \emph{Destination} rule or not). Simulation results rejected the passive movement assumption after comparisons between simulated and observed paths of Cpaaa. Such results indicated that target site specification (the \emph{Destination} rule), as a simplified representation of morphogen gradient, is an effective approach for cell migration path learning, especially when regulatory mechanisms lag data collection. The \emph{left-right asymmetry rearrangement} case demonstrated that the framework has the capability to generalize the \emph{Destination} rule to more specific mechanisms (a leader-follower mechanism in this case) to explain certain cell movement behaviors. By comparing simulated cell migration path regulated by the proposed assumptions and the observed path in a reverse engineering perspective, this framework can be used for facilitating new hypotheses during certain developmental processes not only in \emph{C. elegans}, but in other tissues/organisms as well.

This model captures the main aspects of cell movement and provides a new idea that represents cell behaviors with neural networks trained by deep reinforcement learning algorithms. More powerful models can be implemented in the following aspects: (1) Multi-agent reinforcement learning \cite{busoniu2008comprehensive,tampuu2017multiagent} can be used for studying cooperative/competitive cell behaviors by manipulating the rewards in the framework. Such an extension can provide further biological insights. For example, for the {\em Cpaaa intercalation} case, we may investigate whether the certain group of cells (i.e., Cpaaa and its neighbors) works cooperatively (as a result of the intercalation of Cpaaa) or its neighbors actually act competitively with their own rules (but the regulatory rule of Cpaaa is over-dominant). More specifically, we observed that during the last few minutes of the process, the cell ABarpaapp moves to the posterior to become a neighbor of Cpaaa. It is interesting to study whether ABarpaapp helps Cpaaa to intercalate towards the anterior (cooperative behavior, give both cells rewards when the intercalation of Cpaaa is achieved.), or such a migration of ABarpaapp is just due to its dislocation (competitive behavior, ABarpaapp will not be rewarded when Cpaaa achieves the intercalation.). (2) The hierarchical regulatory mechanism is another area of interest. Although the \emph{Destination} rule provides a simplified representation of morphogen gradient, it can be generalized with the formation of certain cell neighbor relationships. In the \emph{Cpaaa intercalation} case, the intelligent cell experiences a series of changes of neighbor relationships before reaching the target site. It is worth investigating whether these relationships play as significant sub-goals to serve the ultimate goal. As presented in \cite{mnih2015human}, the deep Q-network performs poorly on hierarchical tasks. Such tasks require more advanced strategies that are obtained by prior knowledge, which can hardly be represented by the input state. Therefore, future work is immediately needed to implement hierarchical deep reinforcement learning architectures to meet such demands \cite{kulkarni2016hierarchical}. (3) Other advanced training strategies and reinforcement learning algorithms are also worth investigating to improve the performance of the model, such as learning rate decay \cite{zeiler2012adadelta}, continuous control \cite{lillicrap2015continuous}, and asynchronous methods \cite{mnih2016asynchronous}. (4) Finally, we hope to incorporate more biological domain knowledge in the model to simulate more complex cell movement behaviors. As one of our previous effort, we have developed a developmental landscape for mutated embryos \cite{du2014novo,du2015regulatory}. The mutated cell fate information from this research can be integrated as part of the input state to study a cell's migration path in a mutant. With fate-related adjustments of the regulatory mechanisms and the reward functions behind them, we can verified/rejected the hypotheses of certain cell movement behaviors in a mutant based on the extent of differences between the simulated path and the observed path. Furthermore, by comparing the simulation and observation paths, we can design more biological experiments for follow-up investigations. Other concepts, such as cell-cell adhesion, as environmental factors (like the {\em Collision} and the {\em Boundary} rule) can also be incorporated to improve the performance of the model.

\section{Conclusion}
In this paper, we successfully developed a cell movement modeling system by integrating deep reinforcement learning with an ABM framework. Our modeling system can learn a cell's optimal path under certain regulatory mechanisms, and thus it can examine hypotheses by comparing the similarities between the simulation cell migration paths and the observation data. These two capabilities, in turn, provide new opportunities to explore the large datasets generated by live imaging.

\section*{Funding}
This study is supported by an NIH research project grants (R01GM097576). Research in the Bao lab is also supported by an NIH center grant to MSKCC (P30CA008748). 

%
%

\end{document}